\documentclass[aps,prd,twocolumn,nofootinbib,showpacs,superscriptaddress,groupedaddress]{revtex4-1}

\usepackage[T1]{fontenc} 

\usepackage{amsmath}
\usepackage{amssymb}
\usepackage{graphicx}
\usepackage{bm}
\usepackage{epsfig}
\usepackage{amsfonts}
\usepackage{color}
\usepackage{mathtools}
\usepackage[section]{placeins}

\newcommand{\be}{\begin{equation}}
\newcommand{\ee}{\end{equation}}

\newcommand{\bea}{\begin{eqnarray}}
\newcommand{\eea}{\end{eqnarray}}
\newcommand{\bean}{\begin{eqnarray*}}
\newcommand{\eean}{\end{eqnarray*}}

\begin{document}

\title {Bayesian inference of non-positive spectral functions in quantum field theory}
\author{Alexander Rothkopf}
\affiliation{Institute for Theoretical Physics, Universit\"at
  Heidelberg, Philosophenweg 12, D-69120 Germany}
\date{\today}

\begin{abstract}
We present the generalization to non positive definite spectral functions of a recently proposed Bayesian deconvolution approach (BR method). The novel prior used here retains many of the beneficial analytic properties of the original method, in particular it allows us to integrate out the hyperparameter $\alpha$ directly. To preserve the underlying axiom of scale invariance, we introduce a second default-model related function, whose role is discussed. Our reconstruction prescription is contrasted with existing direct methods, as well as with an approach where shift functions are introduced to compensate for negative spectral features. A mock spectrum analysis inspired by the study of gluon spectral functions in QCD illustrates the capabilities of this new approach.
\end{abstract}

\maketitle


\flushbottom


\section{Introduction}
 \label{sec:intro}
 
The determination of real-time properties of strongly interacting quantum field theories is a key aspect in many areas of modern theoretical physics. It ranges from computations of transport properties \cite{Nature1} of ultra-cold atomic gases (see e.g.\ \cite{Enss:2012wh,Enss:2012qh}) at very low $T\sim10^{-9}K$ to the screening of color charges in a quark-gluon plasma created in a relativistic heavy-ion collision \cite{Muller:2013dea} at $T\sim10^{12}K$ (see e.g. \cite{Rothkopf:2011db,Burnier:2014ssa,Burnier:2015nsa}). Only if we can compute dynamical observables in Minkowski time, our knowledge about the fundamental laws of physics can be connected to real-world experiments.

Despite tremendous progress in both analytic and numerical computational strategies, the majority of approaches, which treat quantum field theory from first principles are formulated in unphysical Euclidean time or as advocated recently e.g.\ in \cite{Pawlowski:2016} the imaginary frequency domain. In the context of the strong interactions lattice QCD is the prime example. In turn one repeatedly encounters the need to perform analytic continuations of correlation functions in order to access the sought after real-time information.
 
Analytic continuations are often implemented by use of spectral functions. I.e.\ many phenomenologically relevant correlation functions permit a representation as integral convolutions of a purely real (or purely imaginary) spectral function $\rho(\omega)$ over a kernel function $K$. The variable we wish to continue appears only in the kernel, which is analytically known. The Euclidean kernel and the K\"all\'en-Lehmann kernel represent two important examples
\begin{align}
D(\tau)=\int_{-\infty}^{\infty} d\omega \frac{e^{-\omega\tau}\,\rho(\omega)}{1\mp e^{-\beta\omega}} ,\; D(\mu)=T\int_{-\infty}^{\infty} d\omega \frac{\rho(\omega)}{\omega-i\mu}.\label{Eq:CorrSpecRel}
\end{align}
Such spectral representations are employed in two ways. If the spectra are known they allows us to carry out the analytic continuation by simply substituting $\tau\to -it$ or $\mu\to-i\omega$ into $K$. On the other hand they provide the basis of how to extract spectral information from computable correlation functions in the first place, i.e.\ via a deconvolution.

The input for the deconvolution are correlation functions $D$, evaluated on a finite number $N_d$ of points with a non-vanishing uncertainty $\Delta D$. 
\begin{align}
D_i=\sum_{l=1}^{N_\omega} \Delta\omega_l K_{il} \rho_l,\quad i\in[0,N_d]
\end{align}
The task to extract from such a limited set of data a continuous spectral function sampled at $N_\omega\gg N_d$ frequencies is clearly ill-posed. This is apparent if we try to carry out a naive $\chi^2$ fit of the $N_\omega$ parameters $\rho_l$, which would yield an infinite number of degenerate solutions all reproducing $D_i$ within errors. 

In order to give meaning to the inversion task we turn to Bayesian inference \cite{Jarrell:1996,Bishop:2007}, starting out from Bayes theorem:
\begin{align}
P[\rho|D,I]\propto P[D|\rho,I]P[\rho|I]
\end{align}
It tells us that the probability of a candidate spectral function $\rho$ to represent the spectral function of a specific correlator $D$ is given by the product of the likelihood probability $P[D|\rho,I]$ and the prior probability $P[\rho|I]$. The former encodes how probable it is that the correlator data actually arises from the test spectrum, while the latter denotes how compatible this test functions is to prior information we possess about the spectrum. If we assume that the $D_i$ are obtained from a sampling algorithm with a Gaussian distribution then $P[D|\rho]={\rm exp}[L]$, where
\begin{align}
L=\frac{1}{2}\sum_{i,j=1}^{N_d}(D_i-D^\rho_i)C_{ij}^{-1}(D_j-D^\rho_j)
\end{align}
with the covariance matrix $C_{ij}$ amounts to the well known $\chi^2$ functional, which in turn is regulated by the prior $P[\rho|I]={\rm exp}[-S]$. 

To set up a viable reconstruction prescription we need to devise a regulator functional $S$ that both encodes as much as possible relevant prior information, while at the same time interferes as little as possible with the actual information encoded in the data. This is achieved in two ways: on the one hand, the functional form of $S[\rho,m]$ itself favors certain spectra, on the other hand it depends on a function $m(\omega)$, the so called default model, at which it is extremal. I.e.\ by definition, in the absence of data, $m(\omega)$ must represent the correct spectrum. Conventionally one further introduces a hyperparameter $\alpha$ to weight the influence of prior information versus that of the data \cite{Jarrell:1996}. In our approach due to the particular form of the prior $S$ and similar to the BR method \cite{Burnier:2013nla} we will integrate out $\alpha$ 
\begin{align}
P[\rho|D,I,m]\propto P[D|\rho,I]\int_0^{\infty} d\alpha P[\rho | m,\alpha] P[\alpha]\label{Eq:IntOutAlpha}
\end{align}
assuming complete ignorance about its values $P[\alpha]=1$.

After choosing both $S$ and $m$, the optimization of the posterior probability 
\begin{align}
\left.\frac{\delta P[\rho|D,I]}{\delta \rho}\right|_{\rho=\rho^{\rm Bayes}}=0
\end{align}
provides a point estimate of the most probable Bayesian spectrum, given the correlator data and our prior information.

We would like to remind the reader that for any finite $N_d$ and finite $\Delta D$, two different Bayesian prescriptions will in general give differing answers. Some parts of the reconstructed spectrum may already be fixed by the data, then the variation of either $S$ or $m$ will leave them intact. Other spectral features are governed by prior information and can be identified as such by appropriately varying that input. 

Due to Bayes theorem all Bayesian methods, if implemented correctly, must converge to the correct result in the limit $N_d\to\infty$ and $\Delta D \to 0$. How quickly they do so depends both on the kernel present in the problem, as well as the choice of S. In particular we will see that exponentially damped kernels present a major challenge to reaching this "Bayesian continuum limit".

In most circumstances the spectra one wishes to extract are apriori known to be positive or negative definite. In such cases the number of possible solutions to the inversion problem reduces dramatically. On the other hand there are both technical and physical reasons, why we may encounter spectra with mixed contributions. Computing correlators with mixed source and sink operators for use e.g.\ in a generalized eigenvalue problem \cite{Harris} or subtracting the perturbatively computed high frequency behavior during a renormalization procedure \cite{Burnier:2011jq} are two pertinent examples.

On the other hand there exists a whole class of phenomenologically relevant spectral functions, which are known to exhibit positivity violation. In particular the single particle spectra of quarks and gluons in Landau gauge QCD \cite{Alkofer:2000wg,Cornwall:2013zra}. For the latter it has been shown that even at weak coupling the spectrum at high frequency approaches zero from below. These spectra, not only provide by itself vital information about the dynamical structure of QCD but also can function as input for real-time computations of e.g.\ transport properties in QCD or condensed matter physics. Therefore in the following we will eventually relax the requirement of positive definiteness leaving us exposed to the full severity of the inverse problem.

In case that we may assume positive definiteness of the spectrum, several Bayesian approaches have been put forward, the most popular being the  Maximum Entropy Method \cite{Asakawa:2000tr,Jakovac:2006sf,Nickel:2006mm}. Based on arguments from two-dimensional image reconstruction it proposes to use the Shannon-Jaynes entropy 
\begin{align}
S_{\rm SJ}=\int d\omega \big(\rho-m-\rho {\rm log}\big[\frac{\rho}{m}\big]\big)
\end{align}
as prior functional. In its state-of-the art implementation by Bryan one in addition restricts by hand the the dimensionality of the solution space around the default model to $N_d$. This restriction is controversially discussed in the literature and extensions towards using the full search space have been proposed (see e.g. \cite{Rothkopf:2011}). Other approaches, such as Tikhonov regularization \cite{Dudal:2013yva}, if reinterpreted in a Bayesian language amount to an approach similar to historic MEM but instead with a quadratic prior and a vanishing default model. The choice of $m=0$ tells us that this approach does not exploit the positive definiteness of the spectrum and it may hide the fact that implicitly a default model has been chosen.

Recently a novel Bayesian approach has been developed (BR method \cite{Burnier:2013nla}), which is specifically geared towards the one-dimensional reconstruction problem following from Eq.\eqref{Eq:CorrSpecRel}. Similar to the MEM it is based on four axioms. It  attempts to describe a prior that penalizes deviations from the default model as weakly as possible, while still providing a unique global answer
\begin{align}
S_{\rm BR}=\int d\omega \big( 1- \frac{\rho(\omega)}{m(\omega)} + {\rm log}\big[ \frac{\rho(\omega)}{m(\omega)} \big]\big).
\end{align}
While it appears to be closely related to the Shannon-Jaynes entropy, it differs in essential ways. Note that only ratios of $\rho$ and $m$ contribute and that the logarithm is not multiplied with the spectrum itself. The latter avoids the problem of flat directions inherent in the Shannon-Jaynes entropy if $\rho/m\ll 1$ and allows us to easily compute the normalization of the corresponding prior probability $Z_S=\int d\rho {\rm exp}[-\alpha S]$.

Now let us have a look at existing strategies on how to accommodate spectra with negative contributions in a Bayesian fashion. The most straight forward is the case of the quadratic prior used e.g.\ in the Tikhonov approach \cite{Dudal:2013yva}, which from the outset permits such spectra. One needs to keep in mind however that this choice of prior leads to a relatively strong penalty of deviations from the default model, which in practice leads to significant smoothing effects that need to be compensated for by an increased number of provided datapoints.

The MEM has been generalized \cite{Hobson:1998bz} by assuming a decomposition of the spectrum into a positive and negative definite part $\rho=\rho_+ - \rho_-$, for each of which a separate prior probability of Shannon-Jaynes type is introduced. This requires the specification of two default models $m_+$ and $m_-$, which together can be combined in a single regulator functional
\begin{align}
S_{+-}[\rho,m_+,m_-]=\int d\omega \big( \psi-m_+-m_--\rho{\rm log}\big[ \frac{\psi+\rho}{2m_+} \big]\big)
\end{align}
with the abbreviation $\psi=\sqrt{\rho^2+m_+m_-}$. While the derivation based on multiplication of probabilities is straight forward, the requirement to provide an apriori decomposition of the spectrum into positive and negative parts is challenging, if not impractical. Often the point of crossing from the positive to the negative domain is information we wish to obtain by carrying out the spectral reconstruction in the first place. In addition we will see that also $S_{+-}$, similar to $S_{\rm SJ}$, leads to a stronger penalty for deviations from the default model, compared to the generalized BR method we will introduce here. 

Let us note that due to the linearity of integrals we may also attempt to circumvent the issue of negative spectra by adding to the correlator datapoints, which arise from a spectrum that only contains positive contributions. The idea is to overcompensate any possible negative spectral contributions in the original dataset. In Euclidean time e.g.\ the convexity of correlators is intimately connected with spectral positivity. I.e one can always construct a shift function with a large enough amplitude so that convexity is restored in the sum of data and shift. We will see that while this approach is straight forward in principle, in practice the dependence on the choice of shift function introduces uncertainties that require a large number of datapoints to be controlled.

With the goal to devise a Bayesian reconstruction for general spectra that does not require an explicit decomposition into positive and negative domains and which leaves as much as possible freedom for the information in the data to manifest itself in the reconstructed spectrum, we introduce in the following section \ref{sec:BRgen} a generalization of the BR method. Sec.\ref{sec:Mock} provides a detailed mock data analysis for the reconstruction of spectra inspired by gluon spectral functions in QCD before concluding in Sec.\ref{sec:Conclusion}

\section{The generalized BR prior}
\label{sec:BRgen}

The starting point of the original BR method are four axioms, two of which are specifically tailored to the one-dimensional reconstruction of positive definite spectra. The first of these is scale invariance, which is related to the fact that by definition the default model must possess the same dimensionality or units as the reconstructed spectrum. It leads one to consider only ratios of $\rho/m$ in the construction of the prior. 

Now that we allow negative contributions in $\rho$ and $m$ and in particular $m=0$ is possible, we may not form a simple ratio of these two quantities. If we wish to determine the deviation of the spectrum from the default model in this case, we can use the difference instead $\rho-m$. Since this quantity is not dimensionless it would not satisfy scale invariance and thus needs to be amended by an additional function which carries the same dimensions as both $\rho$ and $m$. Just as in the case of the likelihood, where the covariance matrix took on a similar role, we may introduce a generalized default model $h(\omega)$, which encodes how confident we are in our default model. This leads us to the ratio $(\rho-m)/h$ that satisfies scale invariance.

The second of the axioms in the original BR method is related to a smoothness assumption, designed to constrain the reconstruction uniquely, where data alone is not able to do so. It is introduced by comparing the values of the ratio $\rho/m$ at neighboring frequencies. Since we wish to use the same path of deriving the analytic form of the generalized prior from a similar argument, we construct a suitable substitute for the ratio. Our ansatz is $|\rho-m|/h+1$, since it measures a weighted deviation between $\rho$ and $m$ with minimal value unity. Following the smoothness argument of the original BR method we thus end up with the following generalized prior function
\begin{align}
&S_{\rm BR}^g[\rho,m,h]=\int d\omega s_{\rm BR}^g[\rho,m,h]=\\
\nonumber&\int d\omega \Big( - \frac{|\rho(\omega)-m(\omega)|}{h(\omega)} + {\rm log}\Big[ \frac{|\rho(\omega)-m(\omega)|}{h(\omega)}+1 \Big]\Big).
\end{align}
\begin{figure}[b]
\includegraphics[scale=0.305, trim = 0.5cm 0.2cm 0.5cm 0.cm, clip=true]{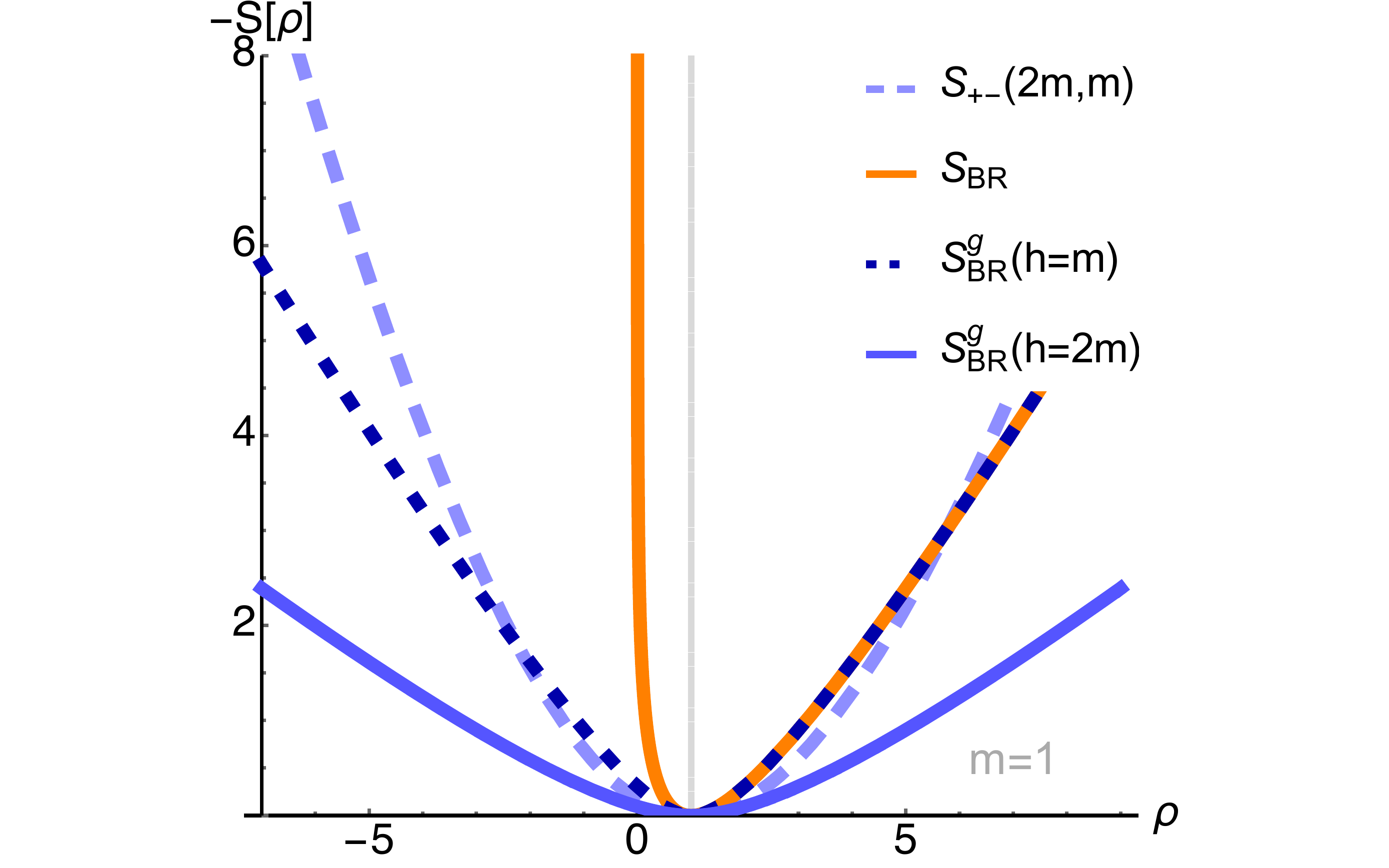}
\caption{Shapes of (minus) the original BR method prior integrand $s_{\rm BR}$ (orange solid) and its extension to non-positive spectra $s^g_{\rm BR}$ for $m=1$. The latter is plotted for two different values of the generalized default model $h=m,2m$ (blue dashed, solid). As comparison we show the generalized MEM prior $s_{+-}(2m,2)$ for a particular positive-negative decomposition (light blue long dash).}\label{Fig:Priors}
\end{figure}
One might worry that the appearance of an absolute value in the integrand will lead to non-differentiability at the extremum $\rho=m$, however both the first and second derivative are well defined everywhere. In particular $\partial_\rho^2 s_{\rm BR}^g>0$, which means that the proof of the existence of a unique maximum of $P[\rho|D,I]$ and thus a unique Bayesian spectral reconstruction still holds, as discussed in \cite{Asakawa:2000tr}. The normalization for $P[\rho|I]=e^{\alpha S^g_{\rm BR}}/Z_S$ reads
\begin{align}
Z_S&=\int {\rm exp}[\alpha S^g_{\rm BR}]\\
\nonumber&= \prod_l 2 h_l e^{\Delta\omega_l \alpha} (\Delta\omega_l\alpha)^{-1-\Delta\omega_l\alpha} \Gamma(1+\Delta\omega_l\alpha,\Delta\omega_l\alpha),
\end{align}
where $\Gamma(a,z)=\int_z^\infty t^{a-1}e^{-t}dt$ denotes the incomplete gamma function.

Let us see how this functional compares to other regulators. To this end we plot in Fig.~\ref{Fig:Priors} for $m=1$ the $\rho$ dependence of minus the integrand $s$ of the prior $S$ at a single frequency.  Besides the original BR method (orange solid), a generalization of the MEM based on the apriori positive-negative decomposition $s_{+-}(2,1)$ (light blue long dashed) is shown. The generalized BR prior for two choices of the generalized default model $h=1,2$ corresponds to the blue dashed lines. As required by Bayes theorem, all curves show an extremum at $\rho=m$ (gray vertical line). $s_{\rm BR}$ forbids negative spectral components, thus it diverges at $\rho=0$. 

If $h=m$ and $m>0$ we have the special case that $s_{\rm BR}$ and $s^g_{\rm BR}$ coincide for $\rho>m$, one essentially mirrors the left hand side of the original BR prior. In the more general case that $h\neq m$ we find 
\begin{align}
&\frac{d^2}{d\rho^2}  s^g_{\rm BR}[\rho,m,h] = \frac{1}{(h+|\rho-m|)^2}, \\
&\frac{d^2}{d\rho^2} s_{+-}[\rho,m_+,m_-] = \frac{1}{\sqrt{m_1 m_2 + \rho^2}},
\end{align}
telling us that away from the extremum the new generalized prior $s^g_{\rm BR}$ shows a weaker curvature than  $s_{+-}$, while still rising faster than linear in $\rho$. 

This fact is important as it means that among the available approaches for arbitrary spectra it imprints the prior information encoded in $m$ most weakly, while still providing a unique answer. In other words it lets the data speak most freely. Conversely this also means that the generalized BR method, just as in the positive definite case is also the most susceptible method to numerical ringing, which needs to be appropriately identified and distinguished from possible peak structures actually encoded in the correlator data $D_i$. In the next section we will find out through a realistic mock data analysis where the benefits and drawbacks of this new prior lie. 

\section{Mock data analysis}
\label{sec:Mock}

\begin{figure}[h]
\includegraphics[scale=0.32]{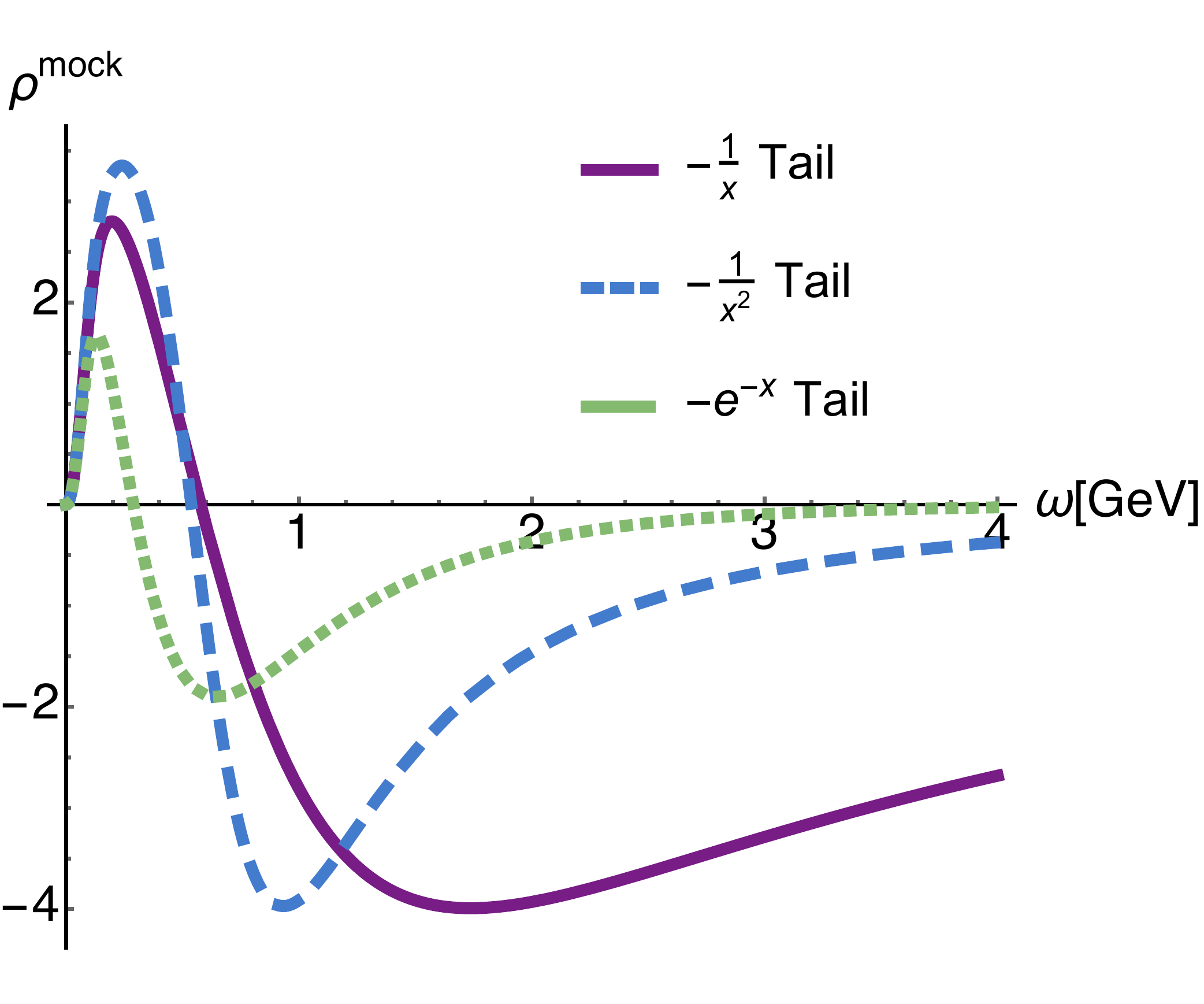}
\caption{Low frequency region of the three mock spectra used. They all contain a positive peak close to the origin, as well as a negative tail structure with either linear (mock1, solid) quadratic (mock2, long dash) or exponential (mock3 , short dash) falloff. The full frequency range covers $\omega\in[0,3000]$GeV to allow the tail to approach the $\omega$-axis appreciably.}\label{Fig:MockSpecs}
\end{figure}

Among others, we plan to deploy this generalized BR method in the future for the study of gluon spectra in Landau gauge QCD using correlation functions obtained from lattice QCD \cite{Ilgenfritz:2017kkp} and the functional renormalization group \cite{Pawlowski:2016inprep}. Therefore we thoroughly test our approach here with mock spectra, which qualitatively resemble the expected functional form of the gluon spectral function. I.e.\ we take these spectra to be anti-symmetric around the frequency origin and as shown in Fig.~\ref{Fig:MockSpecs} they all start out with a quadratic rise and exhibit a peak before entering a negative region. 

\begin{figure}[h]
\includegraphics[scale=0.32]{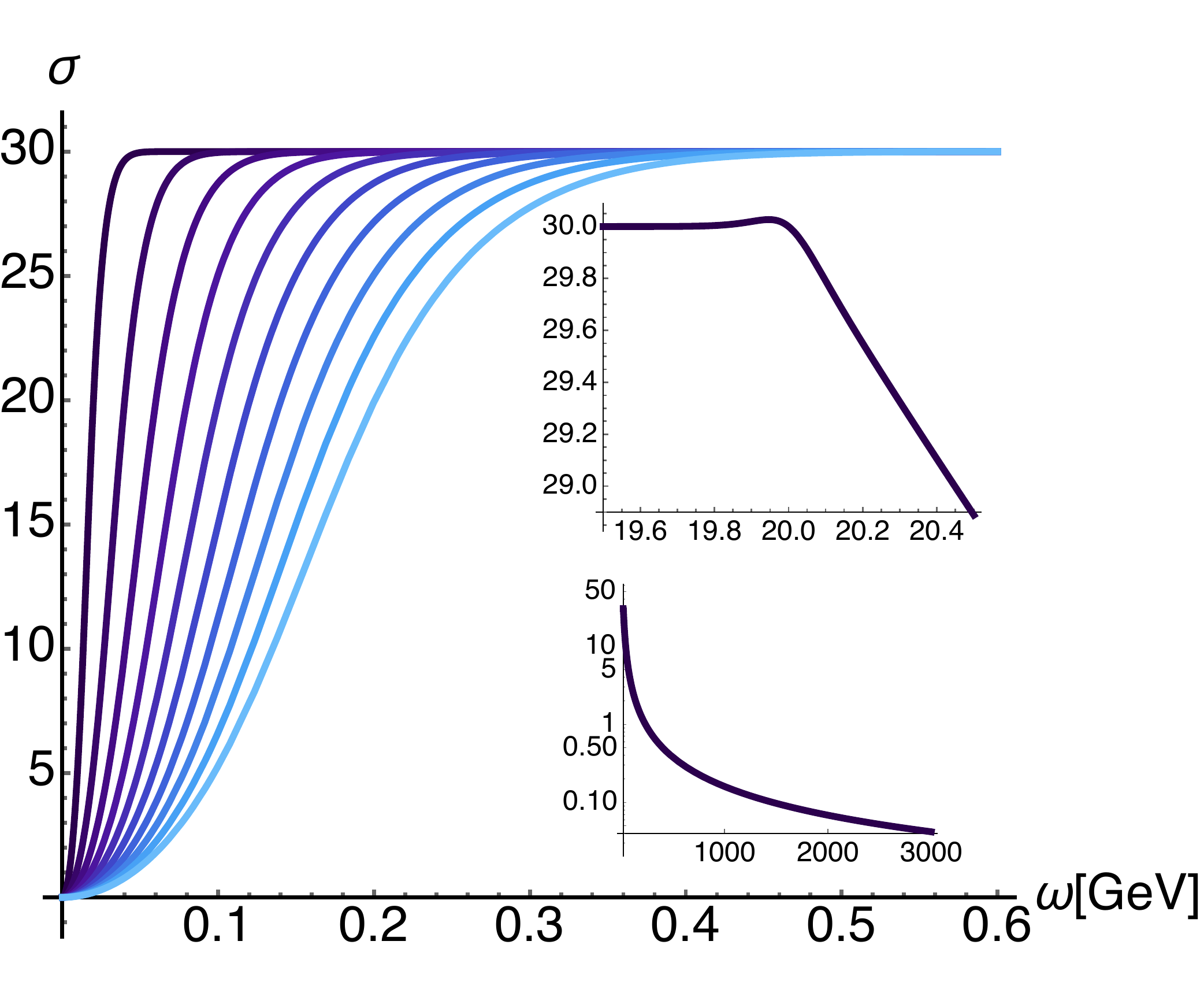}
\caption{The ten different shift functions used in the mock analysis based on the standard positive definite BR method. At small frequencies they exhibit a quadratic rise, which is smoothly cut off at increasing values of $\omega$. At higher frequencies the shift goes over into a  $1/\omega{\rm log}[\omega]^{35/22}$ tail.}\label{Fig:ShiftFuncs}
\end{figure}

For large frequencies the mock spectra approach the $\omega$-axis from below with a tail of different strength. The most difficult (and unphysical) test case is an asymptotic $1/x$ falloff ($\rho^{\rm mock}_1$, solid) the second most challenging amounts to a quadratic behavior ($\rho^{\rm mock}_2$, long dashed) and the easiest one is exponentially damped  ($\rho^{\rm mock}_3$, short dash). To accommodate the tail behavior, the reconstruction is performed on a relatively large frequency interval of $\omega\in[0,3000]$GeV, over which the tails close in on the $\omega$-axis. Such a large frequency range represents an extreme choice, which will not become necessary when e.g. reconstructing from lattice QCD correlators. On the other hand if reconstructions are carried out using correlation functions obtained from continuum computations, weakly damped UV tail structures may be encoded in the data and we wish to ascertain the robustness of our method even in such cases.

We will in the following pit the generalized BR method against another popular approach, where shift data $D^\sigma_i$ are added to the original correlator $D_i$ to overcompensate eventual negative spectral contributions. Subsequently the reconstruction is carried out with a method for positive definite spectra, subtracting in the end the spectrum $\sigma$ corresponding to the shift. Here the original BR method will be deployed to reconstruct the shifted spectrum \footnote{We do not provide a comparison to MEM-related approaches for two reasons. Conceptually the MEM handles both the $\alpha$ integration and the choice of search space differently, which introduces additional systematic uncertainties. On the other hand technically it requires a Singular Value Decomposition, which even with $768$bit precision becomes unstable if such a large frequency range is considered.} using the ten different shift functions given in Fig.~\ref{Fig:ShiftFuncs}.

All of these exhibit a quadratic rise at the origin, which smoothly flattens off around different $\omega_1\in[0.02,0.2]$GeV to a common constant of $s(\omega_1)=30$. At a given $\omega_2=20$GeV $\sigma$ smoothly goes over into a positive falloff, similar in strength to the high frequency perturbative gluon tail. Such a shift fully compensates the negative tail in each of the mock spectra. Note that we will not use any shifts when performing the reconstruction with the generalized BR method later on.

The mock analysis will also allow us to investigate the dependence of the reconstruction on the kernel involved. Since in functional renormalization group computations the correlator can be equally well be evaluated in Euclidean time or imaginary frequency, we will contrast the two cases. Let us start with the more common case of using data along Euclidean time.

\subsection{Reconstruction from Euclidean data}

To generate an ideal set of datapoints we integrate the spectra of Fig.~\ref{Fig:MockSpecs} according to Eq.\ref{Eq:CorrSpecRel} over the exponential kernel. To each $D_i$ we assign an errobar with constant relative magnitude $\Delta D/D={\rm const.}$. Since the Euclidean correlator data is symmetric around $\tau=\beta/2$ it is only necessary to use the first half of the interval, which in turn speeds up the numerics. As in the actual studies, where we will deploy this method, sum rules for the area of the spectrum are available, we replace the first datapoint $\tau=0$ with the integral over the spectrum. This value will be used as an additional constraint on the reconstructed spectrum with equal relative uncertainty.

To obtain the shift data, the functions of Fig.~\ref{Fig:ShiftFuncs} are integrated over as well. We use the {\it Newton-Coates} method in Mathematica, which provides explicit control over the error made. This is particularly important for carrying out the reconstruction from shifted correlators, since the shift data might be one to three orders of magnitude larger than the original data. I.e.\ after adding $D^\sigma_i$ to $D_i$, we also need to appropriately combine their errors, where that of the former may not swamp the relevant signal in the latter. We make sure that $\Delta D^\sigma_i$ is at least an order of magnitude smaller than $\Delta D_i$. In the following the frequency interval is sampled at $N_\omega=4000$ points within three subintervals $[0,2]$GeV with $N_\omega^{(1)}=500$, $(2,60]$GeV with $N_\omega^{(2)}=1500$ and $(60,3000]$GeV with $N_\omega^{(3)}=2000$.

In case of Euclidean time data the integral kernel actually depends on the temperature itself, since the physical length of the imaginary time direction is directly linked to its inverse $\beta$. We test our reconstructions with a kernel corresponding to temperatures between $T=0.1$GeV and $1$GeV.

\subsubsection{Based on shifted data (E)}

\begin{figure*}[t!]
\includegraphics[scale=0.52]{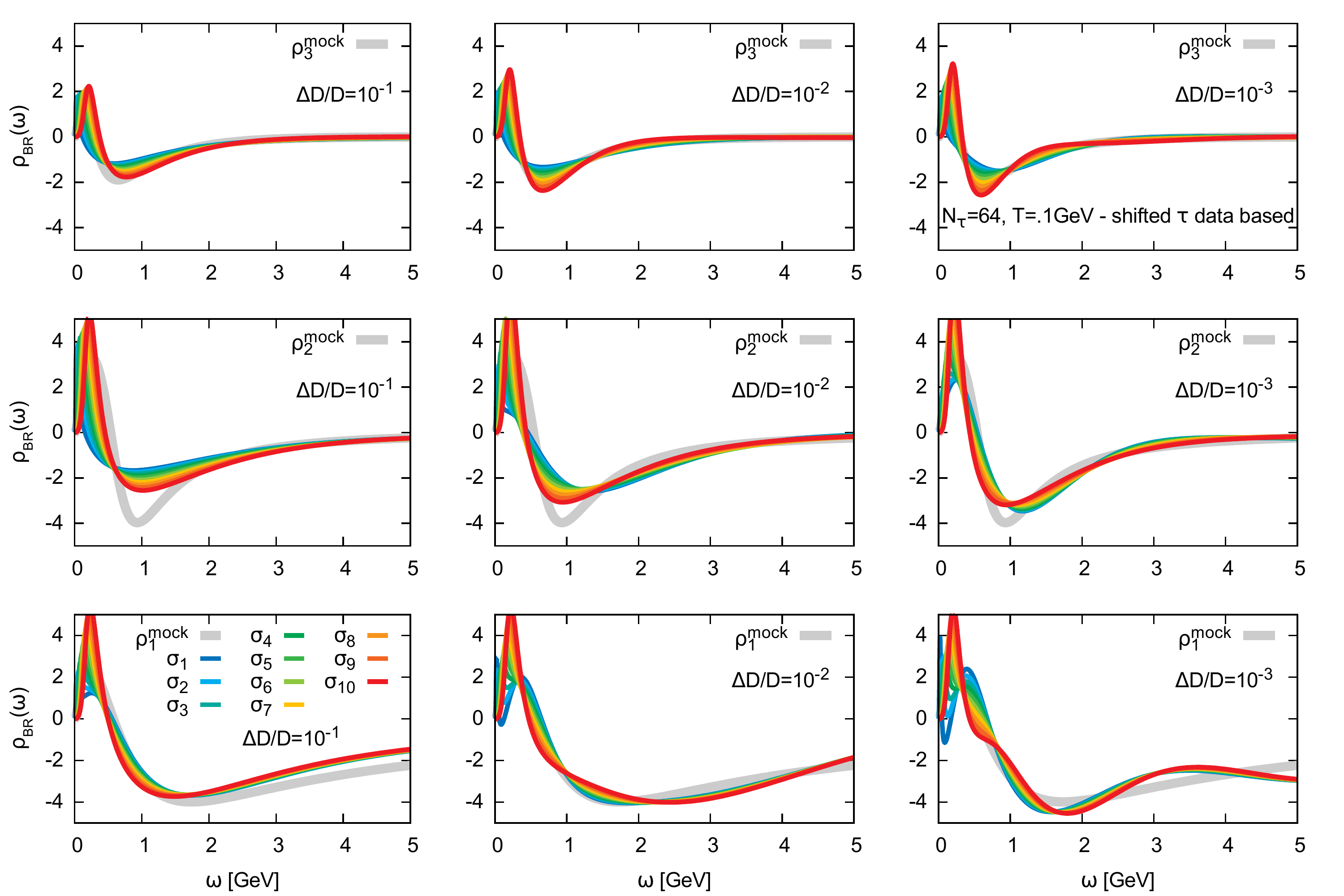}
\caption{Low frequency regime of the spectral reconstructions of mock spectra with asymptotic $-1/x$ ($\rho^{\rm mock}_1$, bottom row),  $-1/x^2$ ($\rho^{\rm mock}_2$, middle row) and exponential tail ($\rho^{\rm mock}_3$, top row) from Euclidean data at $T=0.1$GeV. Only the first half of the symmetric $N_\tau=64$ datapoints is used. The reconstructions are based on shifted data with relative errors $\Delta D/D=10^{-1}$ (left column), $\Delta D/D=10^{-2}$ (center column)  and $\Delta D/D=10^{-3}$ (right column). Each panel shows ten colored curves corresponding to a different shift function $\sigma$ used.}\label{Fig:ALG5EUCLRecN64T01}
\end{figure*}

\begin{figure*}[t]
\includegraphics[scale=0.5]{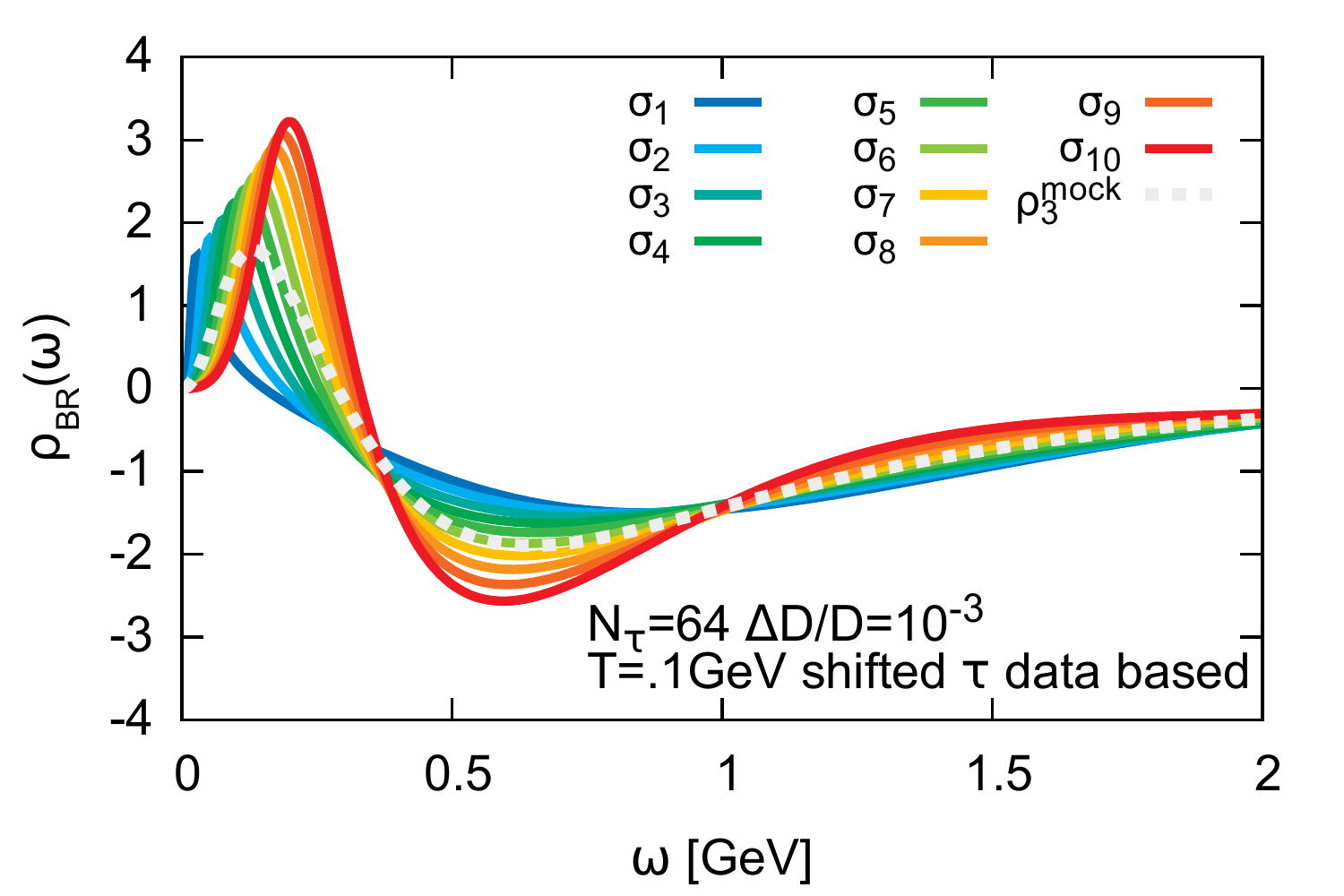}
\includegraphics[scale=0.5]{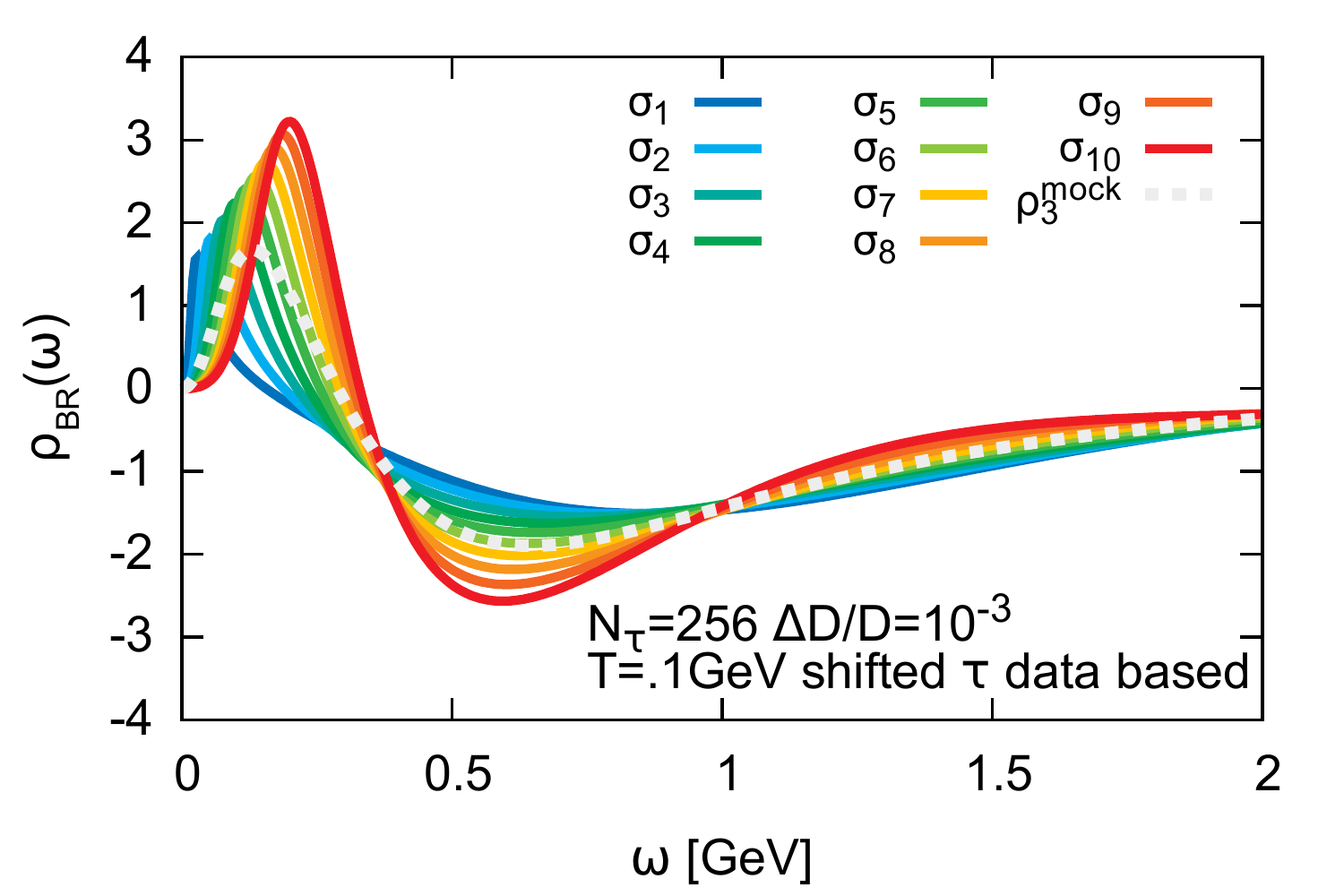}
\caption{Comparison of the very low frequency regime of the reconstructions (colored solid) for the mock spectrum (gray dashed) with exponential tail ($\rho^{\rm mock}_3$) at $\Delta D/D=10^{-3}$ for $N_\tau=64$ (left) and $N_\tau=256$ (right). We find that even a quadrupling of the number of datapoints does not lead to a visible improvement of the reconstruction, in particular the strong dependence on the shift functions remain.}\label{Fig:ALG5EUCLRecN64N256T01}
\end{figure*}

We begin our mock data analysis at the lowest temperature $T=0.1$GeV supplying $N_\tau=64$ shifted datapoints to the original BR method. As default model we take the shift function itself, as we know that it provides a significant contribution to the combined spectrum. The results of the reconstruction are compared for diminishing relative errorbars $\Delta D/D=10^{-1},10^{-2}$ and $10^{-3}$ in the left, center and right column of Fig.~\ref{Fig:ALG5EUCLRecN64T01}. Each row shows the low frequency regime for of one of the three different mock spectra reconstructions. In each panel there are ten curves shown, each representing the outcome for a different shift function.

We find that the reconstruction at this lowest temperature is numerically stable in contrast to previous MEM based analyses \cite{Haas:2013hpa}, where the necessary SVD proved challenging at low $T$.  Even with relatively large errors $\Delta D/D=10^{-1}$ the qualitative structure of the spectrum is reproduced correctly, showing indications of a lowest lying peak, as well as the negative dip and tail. As expected, the overall agreement is best for the mock spectrum with the exponential tail and the largest absolute deviations occur when the asymptotic $-1/x$ tail is considered.

Reducing the errorbars alone only improves the reconstruction in parts of the frequency region, but may at the same time also lead to the appearance of artificial structures in other regions. For the asymptotic $-1/x$ tail (bottom row), decreasing to $\Delta D/D=10^{-3}$ actually emphasizes ringing around the correct result at large frequencies, which depending on the shift function may also influence the small $\omega$ region. 

In the simplest case of the exponential tail mock spectrum (top row) the situation appears better at first sight, with the functional form of the spectrum being reproduced rather well at at $\Delta D/D=10^{-3}$. A closer inspection, as shown in the left panel of Fig.~\ref{Fig:ALG5EUCLRecN64N256T01}, reveals however that we can both significantly diminish the amplitude of the fist peak, as well as the depth of the negative dip by the choice of shift function. In general we find that different $\sigma$'s allow us to easily shift the reconstructed peak position from below to above the actual value. 

Now we may ask whether increasing the number of datapoints will help to remedy the dependence on $\sigma$. While it does so in principle, since due to Bayes theorem in the limit $N_\tau\to\infty$ and $\Delta D/D\to0$ we must recover the correct result, we find that the presence of the exponential kernel hampers us to attain this limit. Indeed, as shown in the right panel of Fig.~\ref{Fig:ALG5EUCLRecN64N256T01}, even if we quadruple the number of datapoints used, there is almost no visible difference in the outcome. In particular the result remains as susceptible to $\sigma$ as before, hinting at the need for an exponentially larger number of datapoints for significant improvement. 

\begin{figure*}[t!]
\includegraphics[scale=0.52]{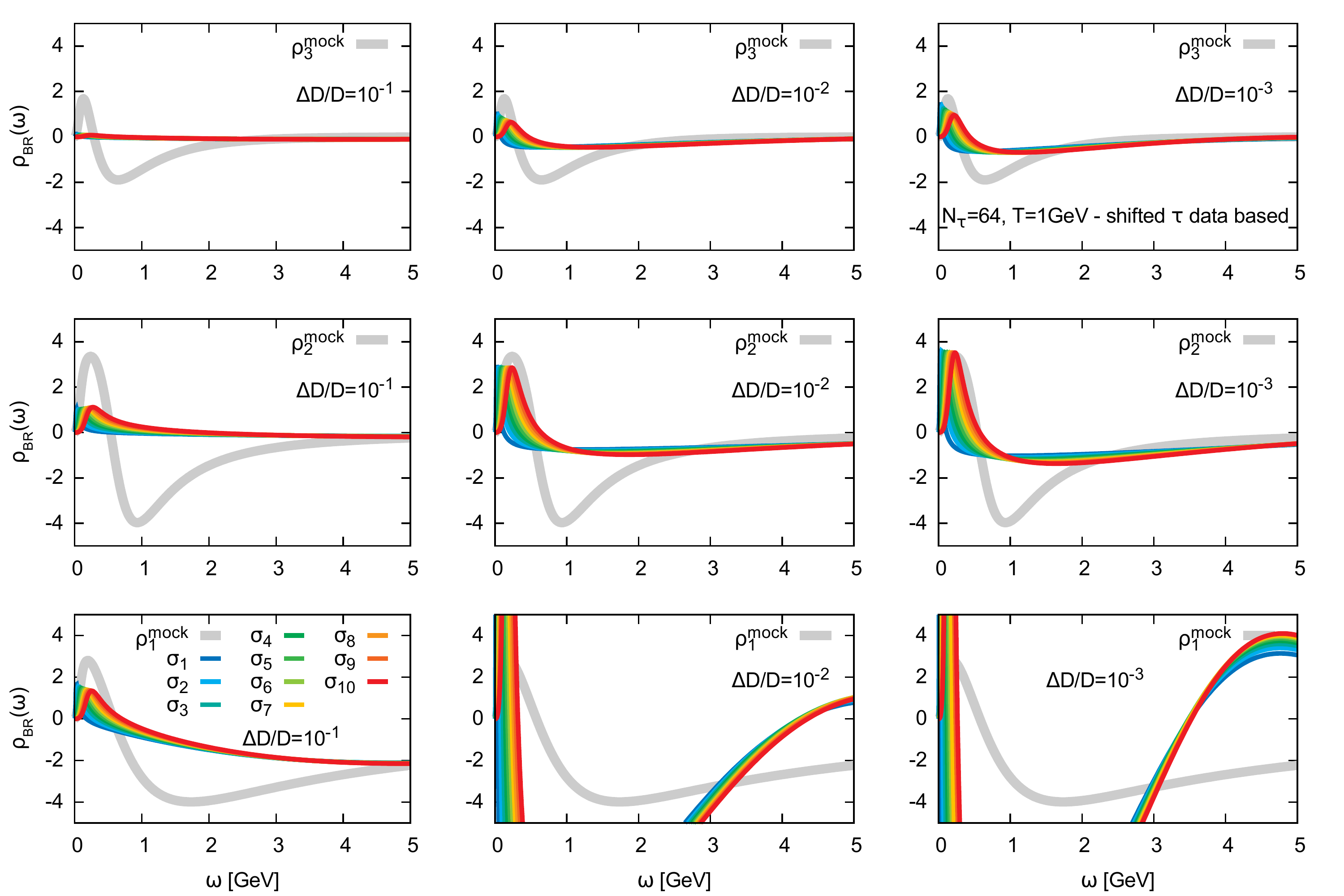}
\caption{Low frequency regime of the spectral reconstructions of mock spectra with asymptotic $-1/x$ ($\rho^{\rm mock}_1$, bottom row), $-1/x^2$ ($\rho^{\rm mock}_2$, middle row) and exponential tail ($\rho^{\rm mock}_3$, top row) from Euclidean data at $T=1$GeV. Only the first half of the symmetric $N_\tau=64$ datapoints is used. The reconstructions are based on shifted data with relative errors $\Delta D/D=10^{-1}$ (left column), $\Delta D/D=10^{-2}$ (center column)  and $\Delta D/D=10^{-3}$ (right column). Each panel shows ten colored curves corresponding to a different shift function $\sigma$ used.}\label{Fig:ALG5EUCLRecN64T10}
\end{figure*}

\begin{figure*}[t]
\includegraphics[scale=0.5]{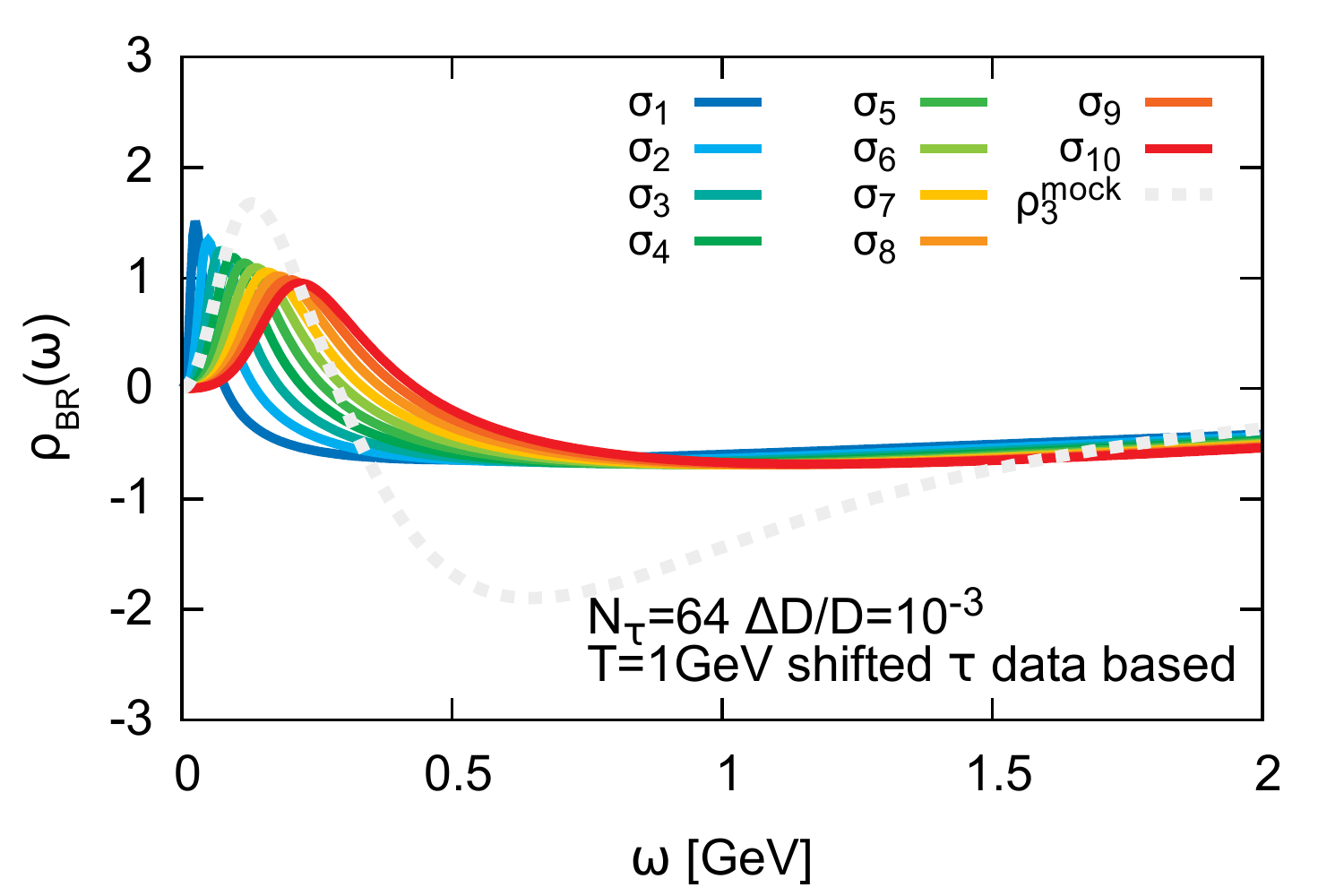}
\includegraphics[scale=0.5]{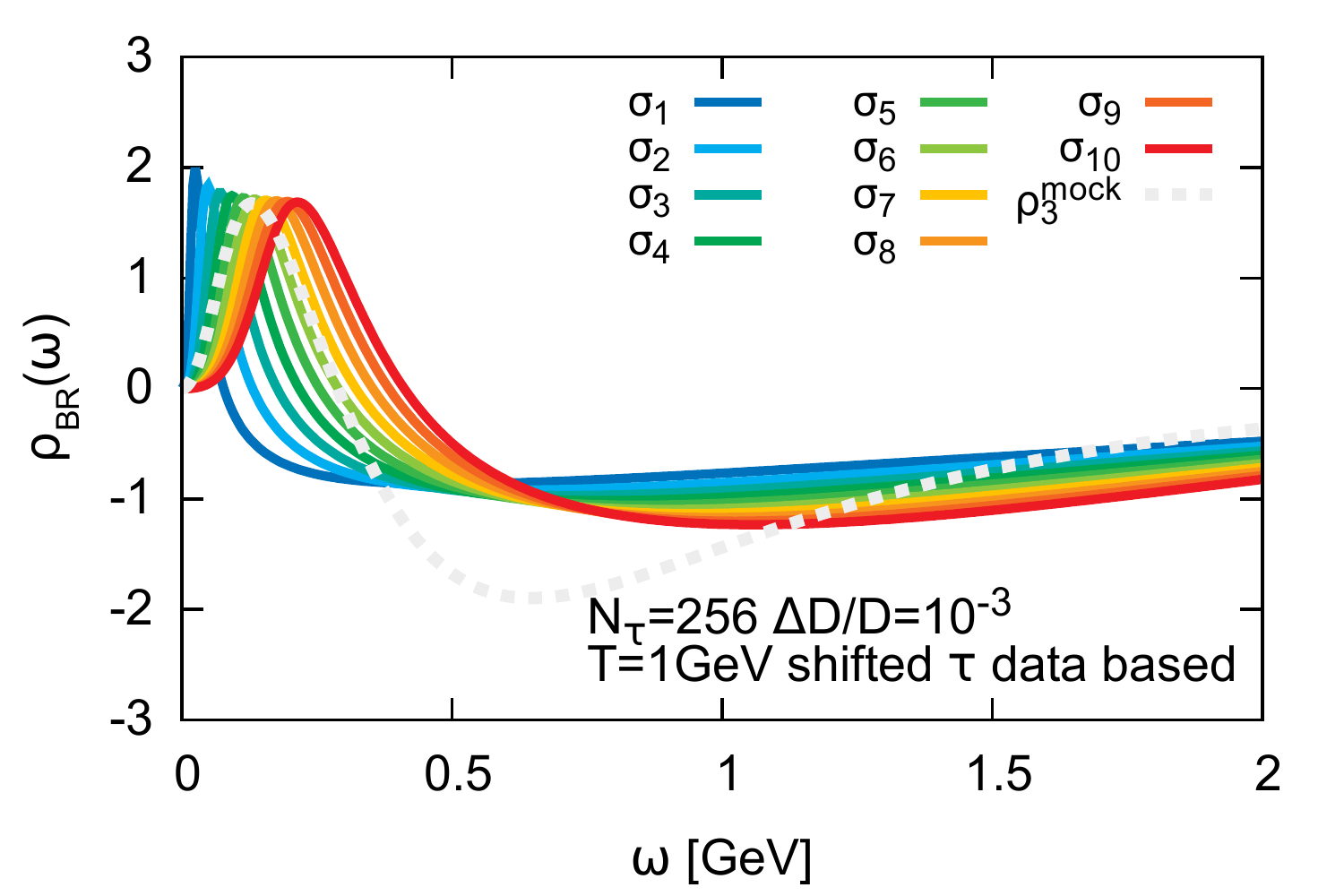}
\caption{Comparison of the very low frequency regime of the reconstructions (colored solid) for the mock spectrum (gray dashed) with exponential tail ($\rho^{\rm mock}_3$) at $\Delta D/D=10^{-3}$ for $N_\tau=64$ (left) and $N_\tau=256$ (right). We find that while quadrupling of the number of datapoints slightly improves the reconstruction of the lowest peak, it does not lead to a significantly better result for the negative trough. In particular the strong dependence on the shift functions remains.}\label{Fig:ALG5EUCLRecN64N256T10}
\end{figure*}

Let us continue by presenting the reconstructions performed on the same number of datapoints, now using a kernel at $T=1.0$GeV as shown in Fig.~\ref{Fig:ALG5EUCLRecN64T10}. The difference to the low $T$ case is striking. For $\Delta D/D=10^{-1}$ we both miss the negative trough completely and only find a shallow enhancement close to the origin, no matter what mock spectrum is chosen. While decreasing the errorbars slightly improves the outcome for $\rho^{\rm mock}_2$ (middle row) and $\rho^{\rm mock}_3$ (top row) it leads to very strong ringing artifacts for $\rho^{\rm mock}_1$ (bottom row). 

If we focus in more detail on the reconstruction of the very low frequency regime shown in the left panel of Fig.~\ref{Fig:ALG5EUCLRecN64N256T10}, we find that indeed even at $\Delta D/D=10^{-3}$ the negative trough for $\rho^{\rm mock}_3$ is barely visible and the position and width of the lowest lying peak dependent strongly on the shift function. Just as at $T=0.1$GeV, without knowledge of the solution it is difficult to choose apriori, which shift function is most appropriate. In particular since choosing a $\sigma$, which leads to a low accuracy for the peak position, does not necessarily induce extra artificial ringing.

\begin{figure*}[t!]
\includegraphics[scale=0.52]{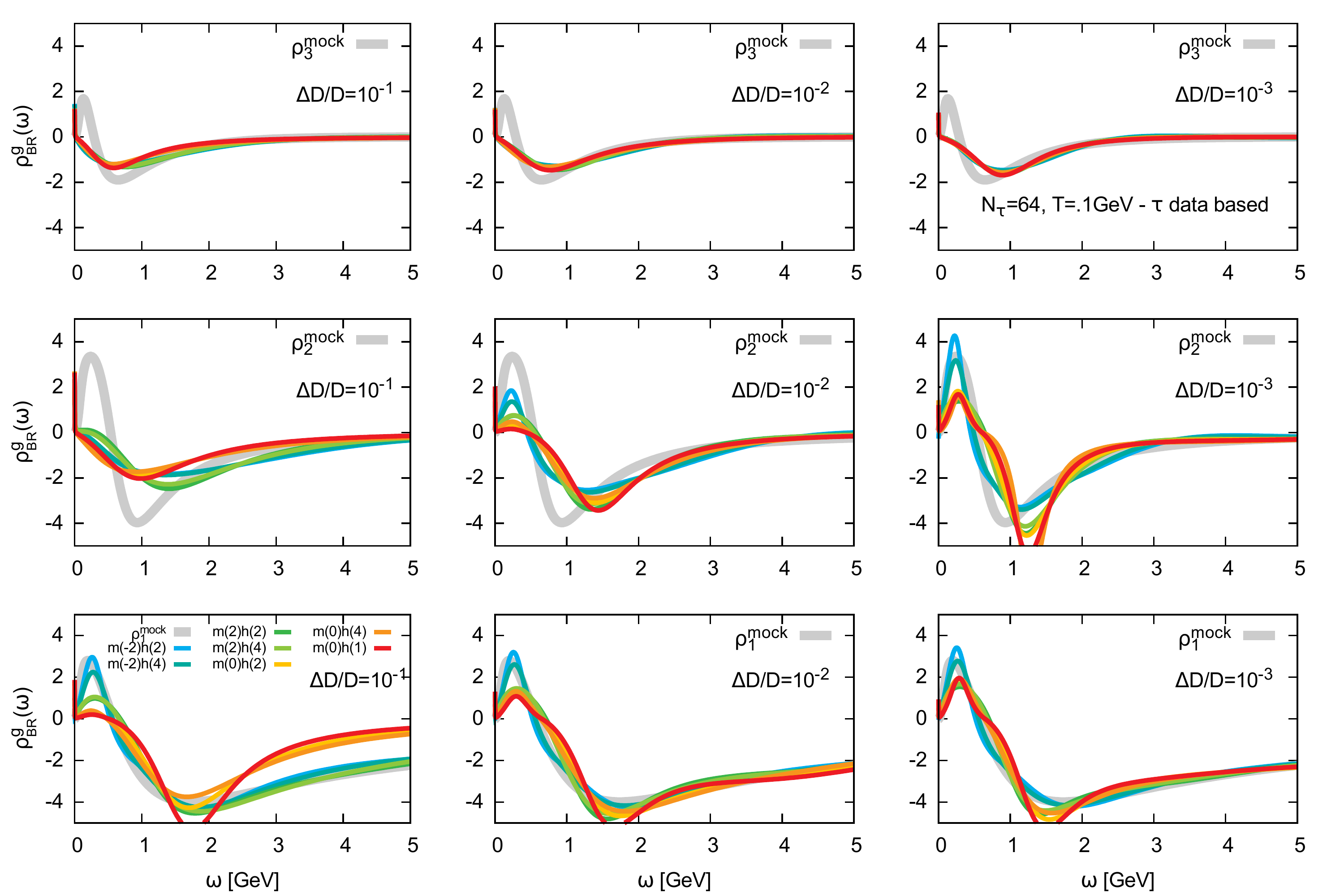}
\caption{Low frequency regime of the spectral reconstructions of mock spectra with asymptotically $-1/x$ ($\rho^{\rm mock}_1$, bottom row), $-1/x^2$ ($\rho^{\rm mock}_2$, middle row) and exponential tail ($\rho^{\rm mock}_3$, top row) from Euclidean data at $T=0.1$GeV. Only the first half of the symmetric $N_\tau=64$ datapoints is used. The reconstructions are based on the generalized BR method using unshifted data with relative errors $\Delta D/D=10^{-1}$ (left column), $\Delta D/D=10^{-2}$ (center column)  and $\Delta D/D=10^{-3}$ (right column). Each panel shows nine colored curves corresponding to different combinations of $m$ and $h$.}\label{Fig:ALG7EUCLRecN64T01}
\end{figure*}

As we may expect from the findings at low temperature, increasing the number of available datapoints by a factor four also at $T=1$GeV does not lead to a significant improvement. As shown in the right panel of Fig.~\ref{Fig:ALG5EUCLRecN64N256T10} going to $N_\tau=256$ leads to a slightly better signal for the lowest lying peak and the existence of the negative trough at least seems to be hinted at in the reconstruction. Nevertheless position and width of the lowest lying peak remain strongly dependent on the choice of shift function $\sigma$. This indication of a very slow approach to the "Bayesian continuum limit" again emphasizes the challenge posed by an exponentially damped kernel.

Note that the systematic uncertainty in the reconstructions above consists not only of the influence of the shift function, but as well increases due to possible different choices of default models. Up to here we have always chosen $m=\sigma$, which we would need to vary to estimate the full error budget. However we have seen that the choice of $\sigma$ alone already leads to significant changes in the position and width of the reconstruction even for high quality Euclidean data $N_\tau=256$ and $\Delta D/D=10^{-3}$. An investigation of different default models is hence superfluous.

With these findings in mind we continue by deploying the novel generalized Bayesian Reconstruction method on unshifted Euclidean data.

\subsubsection{Generalized BR method (E)}

\begin{figure*}[t]
\includegraphics[scale=0.5]{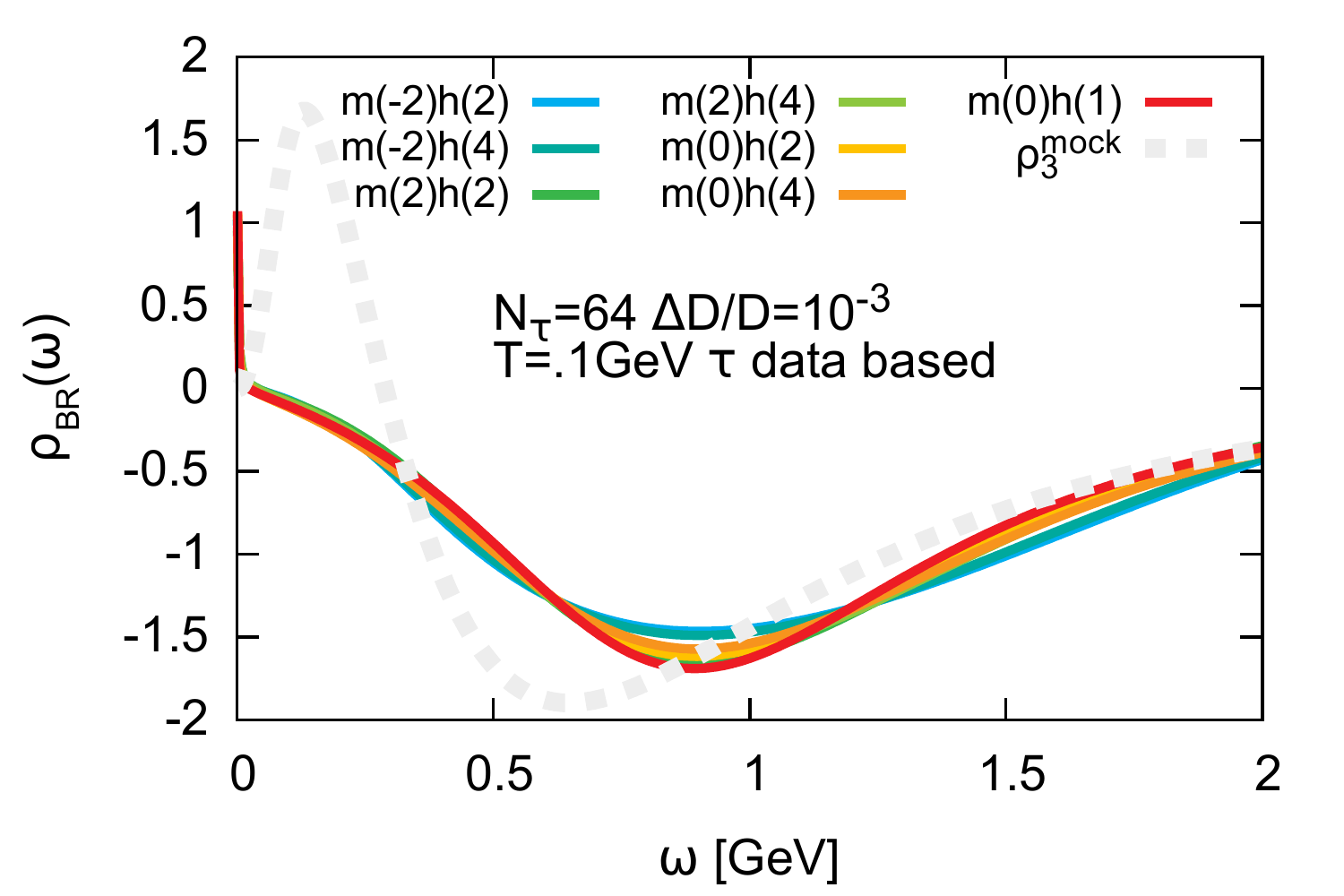}
\includegraphics[scale=0.5]{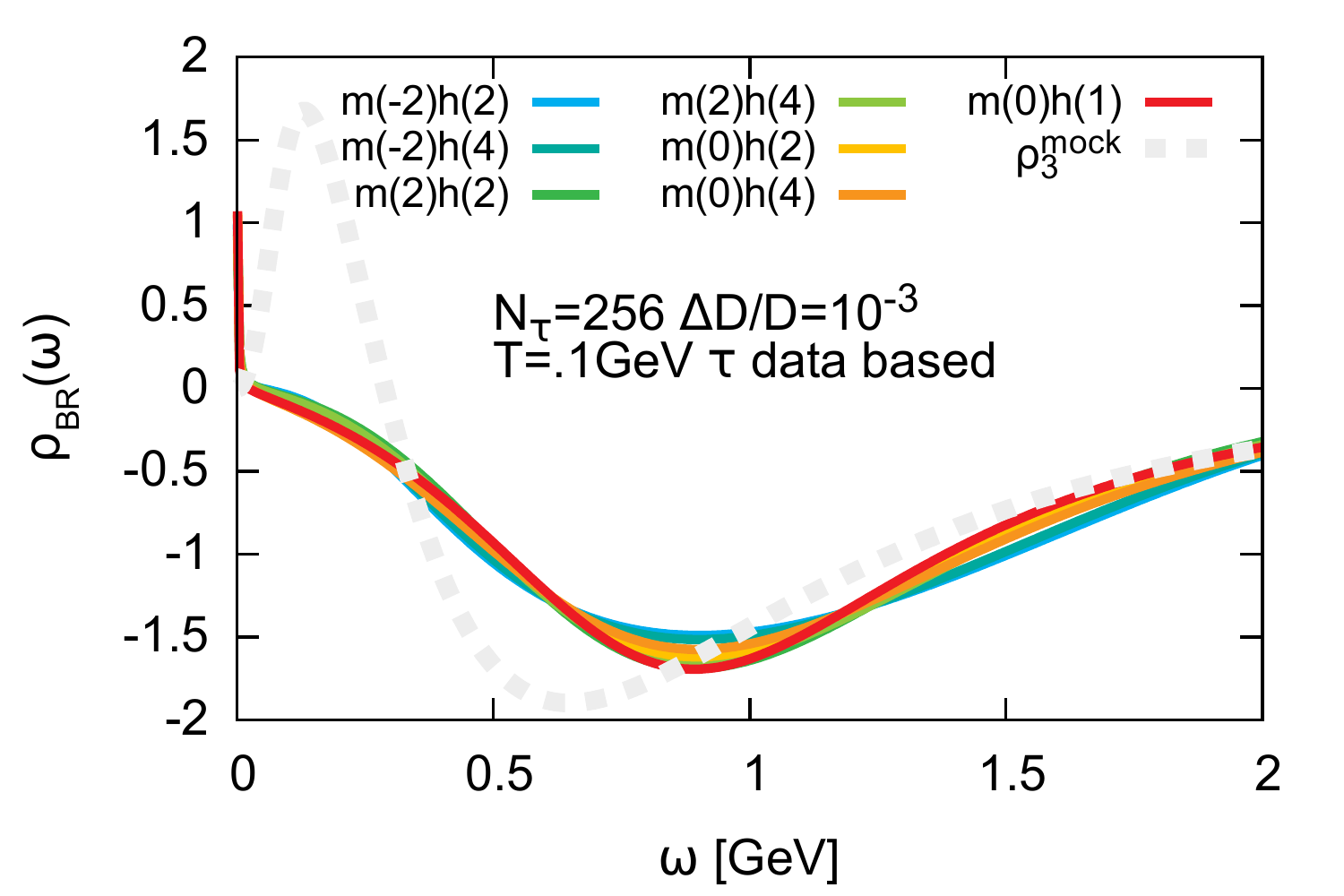}
\caption{Comparison of the very low frequency regime of the reconstructions (colored solid) for the mock spectrum (gray dashed) with exponential tail ($\rho^{\rm mock}_3$) at $\Delta D/D=10^{-3}$ for $N_\tau=64$ (left) and $N_\tau=256$ (right). We find that even a quadrupling of the number of datapoints does not lead to a visible improvement of the reconstruction, in particular the strong dependence on the shift functions remain.}\label{Fig:ALG7EUCLRecN64N256T01}
\end{figure*}

\begin{figure*}[t!]
\includegraphics[scale=0.52]{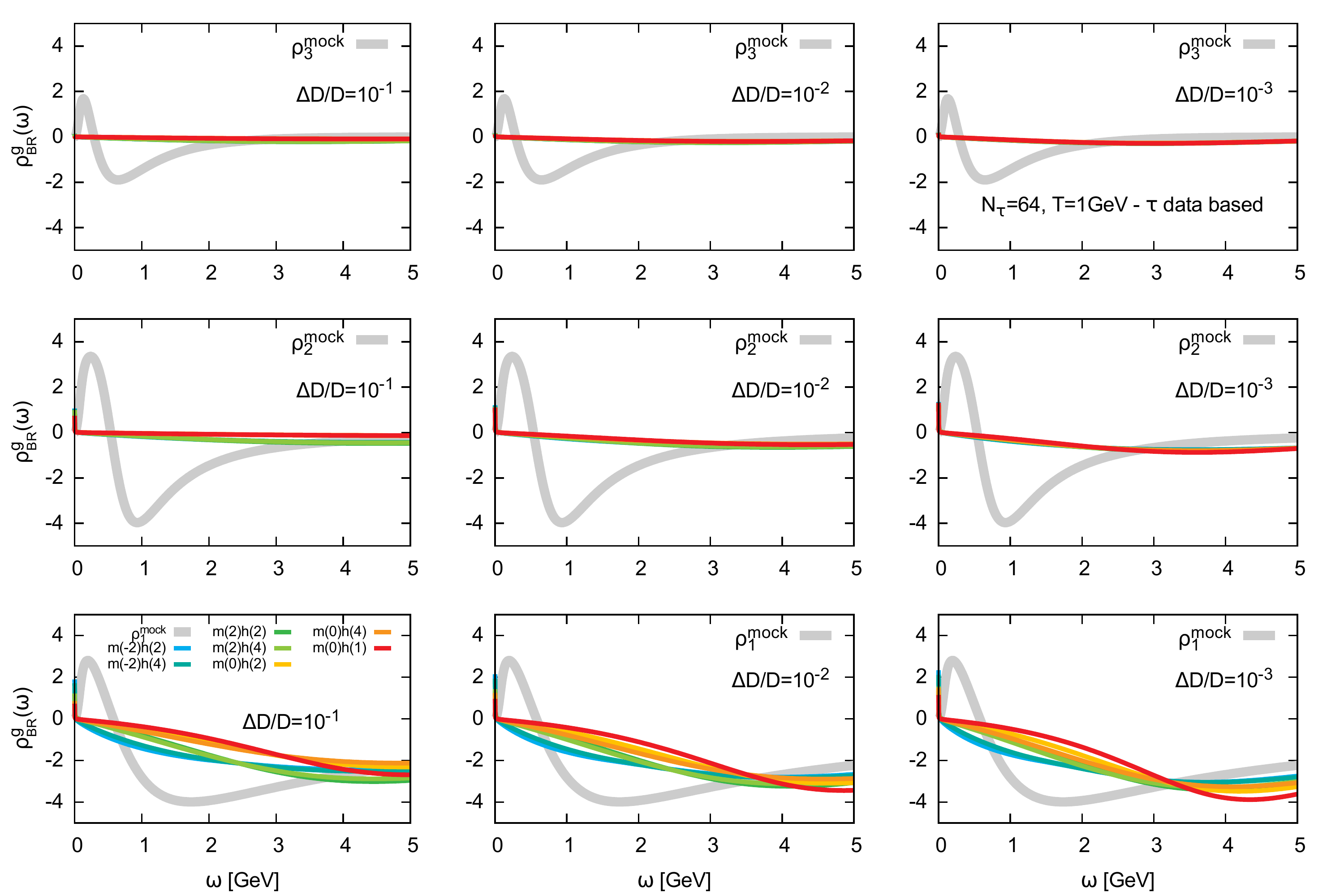}
\caption{Low frequency regime of the spectral reconstructions of mock spectra with asymptotic $-1/x$ ($\rho^{\rm mock}_1$, bottom row), $-1/x^2$ ($\rho^{\rm mock}_2$, middle row) and exponential tail ($\rho^{\rm mock}_3$, top row) from Euclidean data at $T=1$GeV. Only the first half of the symmetric $N_\tau=64$ datapoints is used. The reconstructions are based on the generalized BR method using unshifted data with relative errors $\Delta D/D=10^{-1}$ (left column), $\Delta D/D=10^{-2}$ (center column)  and $\Delta D/D=10^{-3}$ (right column). Each panel shows nine colored curves corresponding to different combinations of $m$ and $h$.}\label{Fig:ALG7EUCLRecN64T10}
\end{figure*}

The generalized BR method does not require us to add a shift contribution to the original ideal Euclidean data and we may choose a default model that also can take arbitrary values, in particular $m=0$ is admissible. On the other hand it asks us to specify a generalized default-model $h$, which may be interpreted as the confidence we have in the values of $m$. While the dependence on a shift function is absent in this approach, we must still elucidate how the choice of $m$ and $h$ influences the outcome. Therefore in the following we will always show nine curves corresponding to the combination of $m=-2,0,2$GeV and $h=2,4$GeV, which gives a rough estimate of the full systematic uncertainties. The reason is that since we integrate out the hyperparameter $\alpha$ in Eq.\eqref{Eq:IntOutAlpha} no further free parameters enter in our generalized approach.

In Fig.~\ref{Fig:ALG7EUCLRecN64T01} we show the outcome of the reconstructions from unshifted $N_\tau=64$ Euclidean data points for $T=0.1$GeV. One may wonder that these curves apparently show more variation than those for the shifted reconstruction in Fig.~\ref{Fig:ALG5EUCLRecN64T01}. But bear in mind that here the different curves correspond to the default model dependence, which we did not take into account before. 

The negative trough is well exposed in these reconstructions, however for e.g.\ $\Delta D/D=10^{-1}$  the information on the lowest lying peak seems to manifest itself only in a very narrow structure close to the origin. While this appears worse than in the case for the shifted reconstruction, note that through the choice of shift function with a quadratic initial rise, we predisposed the shifted reconstruction to show a peaked structure at low frequencies. In that sense the generalized BR method is less biased.

\begin{figure*}[t!]
\includegraphics[scale=0.52]{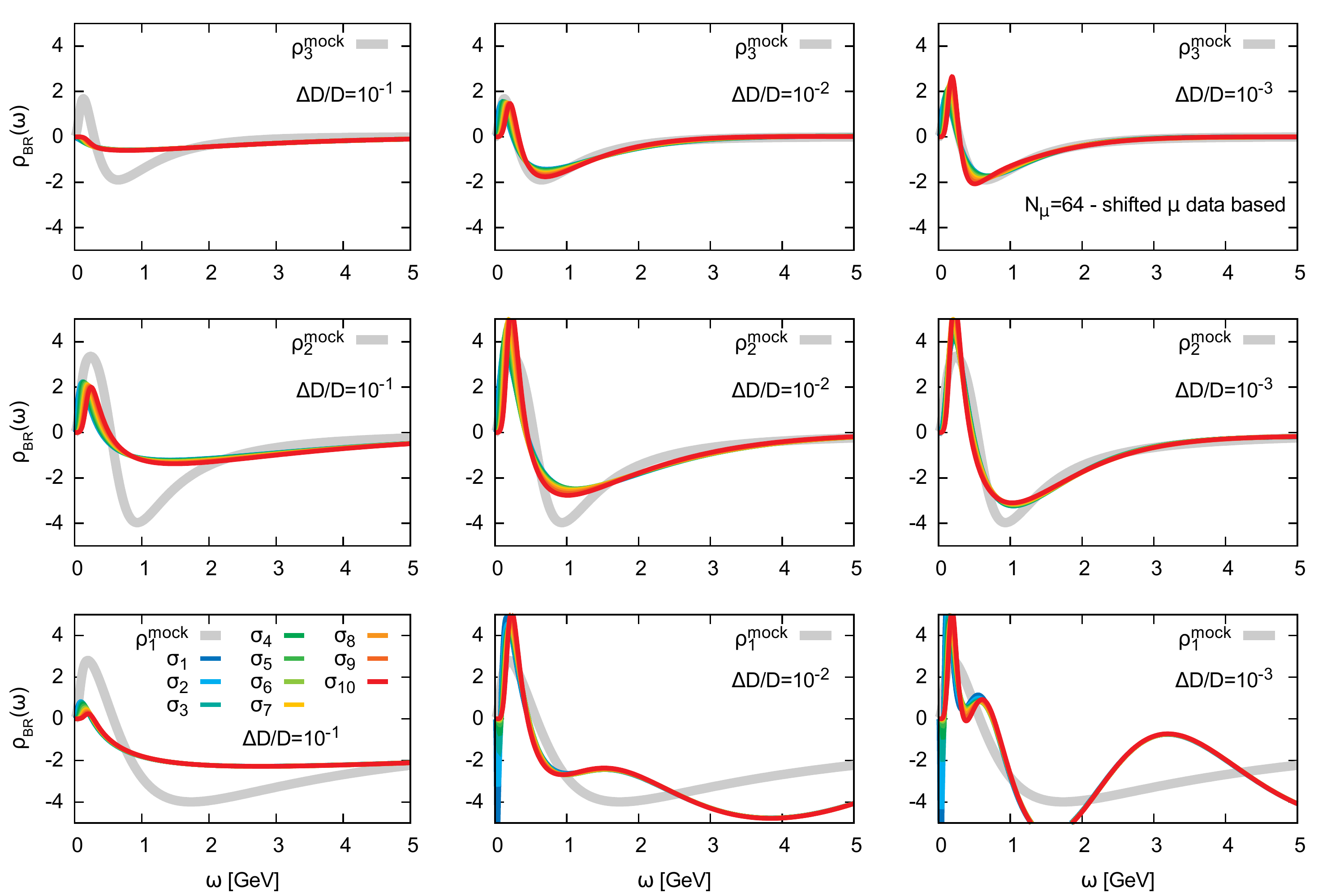}
\caption{Low frequency regime of the spectral reconstructions of mock spectra with asymptotic $-1/x$ ($\rho^{\rm mock}_1$, bottom row), $-1/x^2$ ($\rho^{\rm mock}_2$, middle row) and exponential tail ($\rho^{\rm mock}_3$, top row) from imaginary frequency data. The reconstructions are based on shifted data with relative errors $\Delta D/D=10^{-1}$ (left column), $\Delta D/D=10^{-2}$ (center column)  and $\Delta D/D=10^{-3}$ (right column). Each panel shows ten colored curves corresponding to a different shift function $\sigma$ used. }\label{Fig:ALG5KAELLENRecN64}
\end{figure*}

\begin{figure*}[t!]
\includegraphics[scale=0.5]{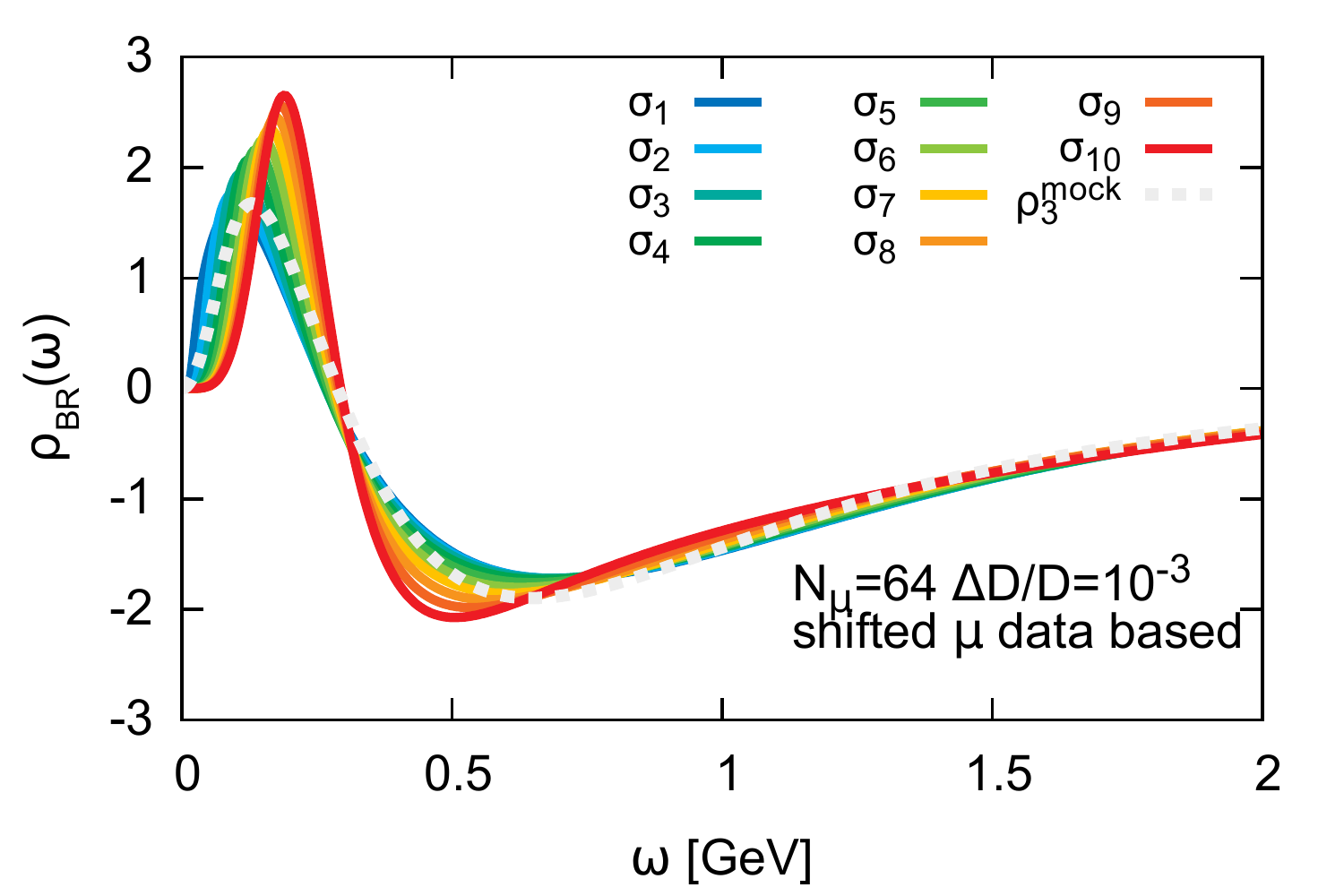}
\includegraphics[scale=0.5]{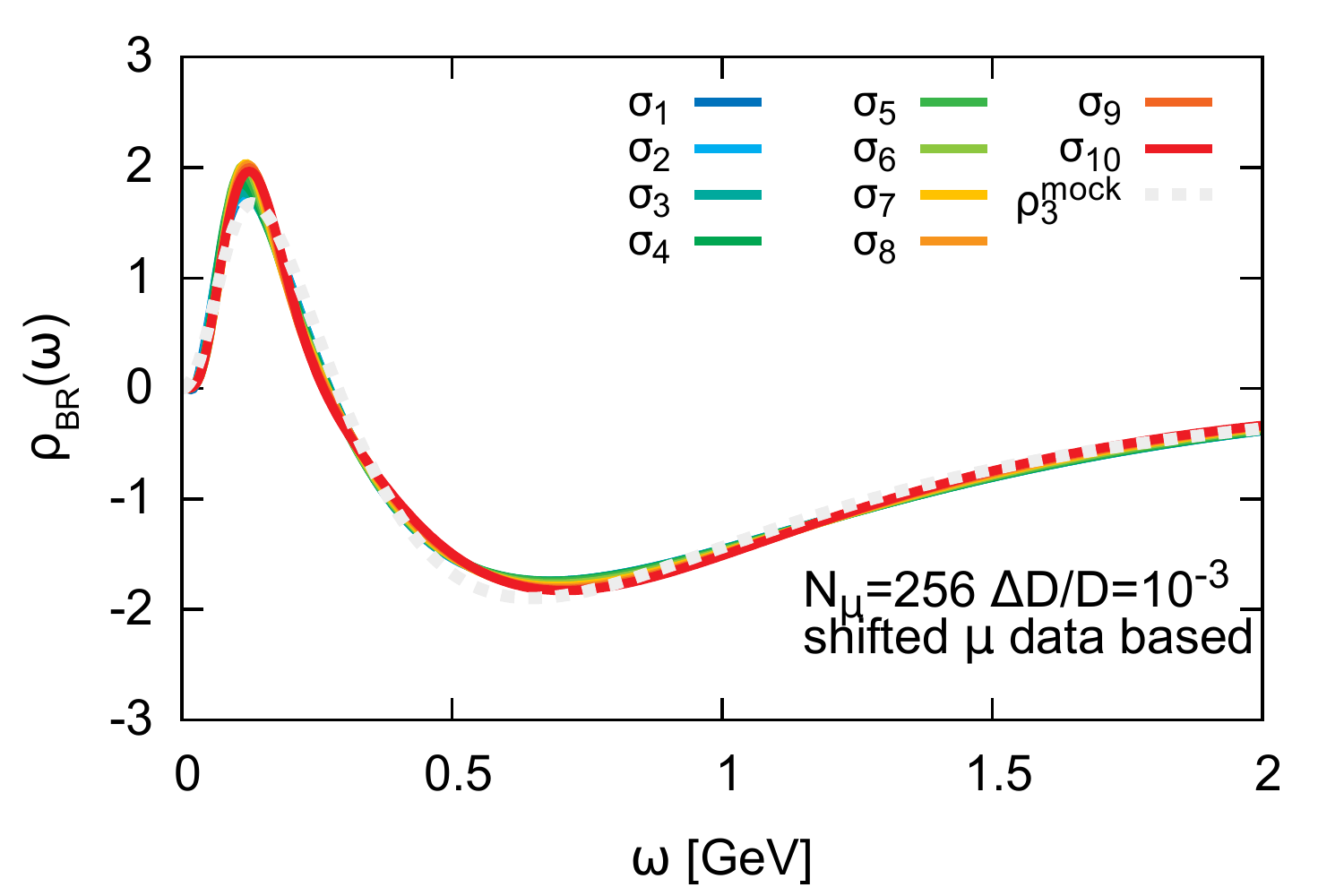}
\caption{Comparison of the very low frequency regime of the reconstructions (colored solid) for the mock spectrum (gray dashed) with exponential tail ($\rho^{\rm mock}_3$) at $\Delta D/D=10^{-3}$ for $N_\mu=64$ (left) and $N_\mu=256$ (right). Contrary to the Euclidean case, increasing the number of imaginary frequency datapoints significantly improves the reconstruction and reduces the dependence on the shift functions $\sigma$.}\label{Fig:ALG5KAELLENRecN64N256}
\end{figure*}

Unfortunately the loss of information due to the exponential Euclidean kernel is independent from the method used to reconstruct the spectrum, which is why as shown in Fig.~\ref{Fig:ALG7EUCLRecN64N256T01}, going over to $N_\tau=256$ also does not improve the reconstruction with the novel generalized approach.

Just as we expect from the shifted analysis, going to higher temperatures leads to an even worse reconstruction result as shown in Fig.~\ref{Fig:ALG7EUCLRecN64T10}. Even with the smallest errorbars $\Delta D/D=10^{-3}$ the lowest lying peak is virtually absent in all reconstructions and the negative trough not even qualitatively captured satisfactorily.

All the mock tests we have performed so far paint a rather bleak picture regarding a successful reconstruction of non-positive spectra from Euclidean datasets. While in principle increasing $N_\tau$ and reducing $\Delta D/D$ will ultimately allow us to arrive at the correct solution, the exponential information loss induced by the Euclidean kernel makes the approach by any practical means prohibitively expensive, especially the higher the temperature is.

We therefore continue by investigating as an alternative the reconstruction performance in case that imaginary frequency data is available. The relation of Eq.\eqref{Eq:CorrSpecRel} with a rational kernel and thus a much weaker information loss bodes well in this regard. In particular the functional approach to QCD to be used to compute data in Ref.~\cite{Pawlowski:2016inprep} allows one to evaluate the correlators directly in the variable $\mu$.

\subsection{Reconstruction from imaginary frequency data}

Similar to the behavior of actual gluon spectra, we consider here anti-symmetric spectral functions, which allows us to rewrite the corresponding relation in Eq.\eqref{Eq:CorrSpecRel} as 
\begin{align}
\int_0^\infty \, d\omega \, \frac{2\omega}{\omega^2+\mu^2+\epsilon}\rho(\omega),
\end{align}
where we regularize the kernel with $\epsilon^2=10^{-15}$GeV$^2$ for consistent limits $\omega\to0$ and $\mu\to0$. If imaginary frequency data is available, the polynomial falloff of the kernel promises a much better reconstruction success, compared to the exponential decay in the Euclidean case. We discretize the imaginary frequency domain $\mu\in[0,8]$GeV with $N_\mu=64$.  The shift functions if used, as well as the choice of default model remains the same as in the Euclidean case. Note that the K\"all\'en-Lehmann kernel does not intrinsically depend on the temperature of the system, so that we only need to consider a single set of reconstructions for a given $N_\mu$ and $\Delta D/ D$ in the following.

\subsubsection{Original BR method (I)}

\begin{figure*}[t!]
\includegraphics[scale=0.52]{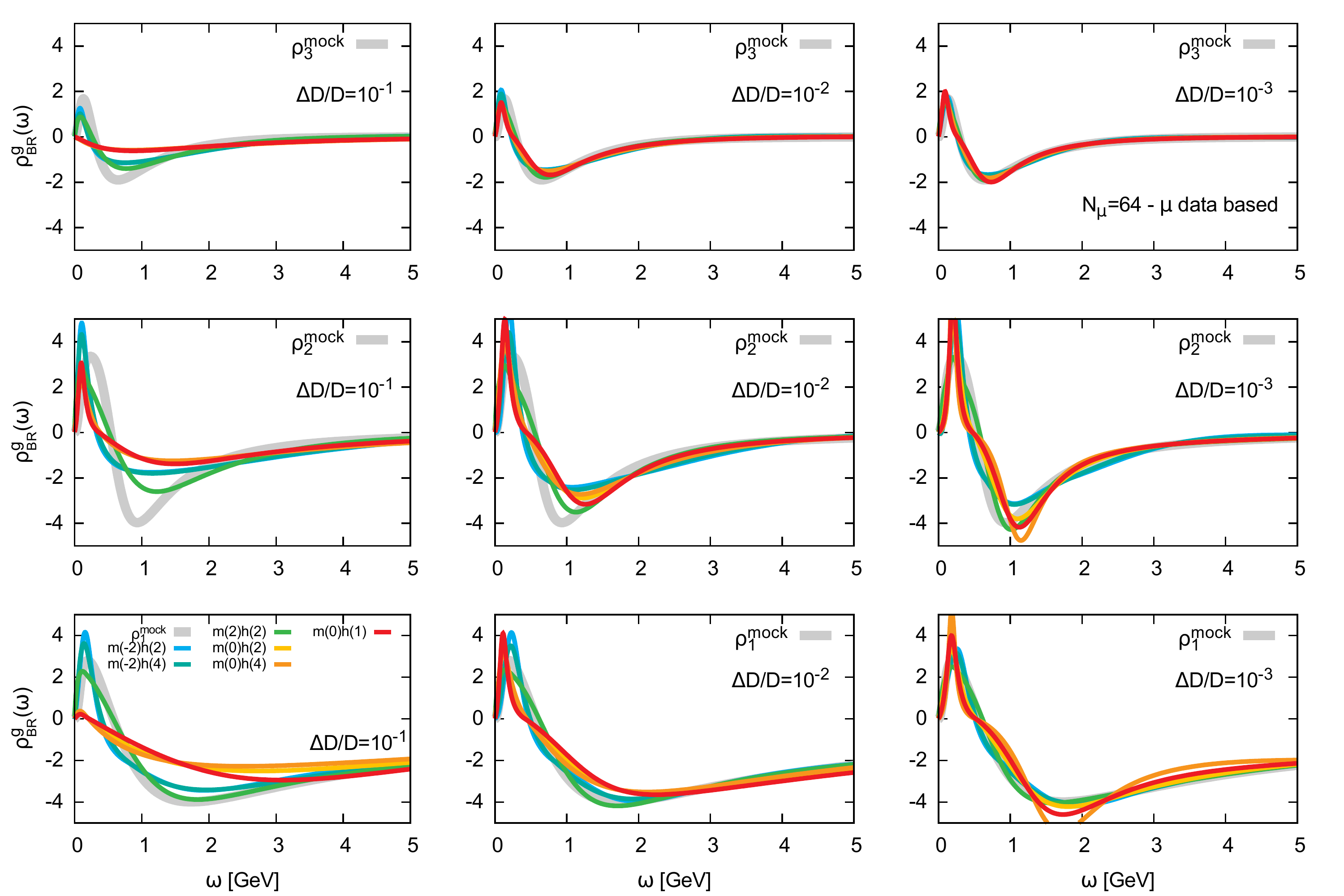}
\caption{Low frequency regime of the spectral reconstructions of mock spectra with asymptotic $-1/x$ ($\rho^{\rm mock}_1$, bottom row), $-1/x^2$ ($\rho^{\rm mock}_2$, middle row) and exponential tail ($\rho^{\rm mock}_3$, top row) from imaginary frequency data. The reconstructions are based on the generalized BR method using unshifted data with relative errors $\Delta D/D=10^{-1}$ (left column), $\Delta D/D=10^{-2}$ (center column)  and $\Delta D/D=10^{-3}$ (right column). Each panel shows nine colored curves corresponding to different combinations of $m$ and $h$.}\label{Fig:ALG7KAELLENRecM1N64}
\end{figure*}

\begin{figure*}[t!]
\includegraphics[scale=0.5]{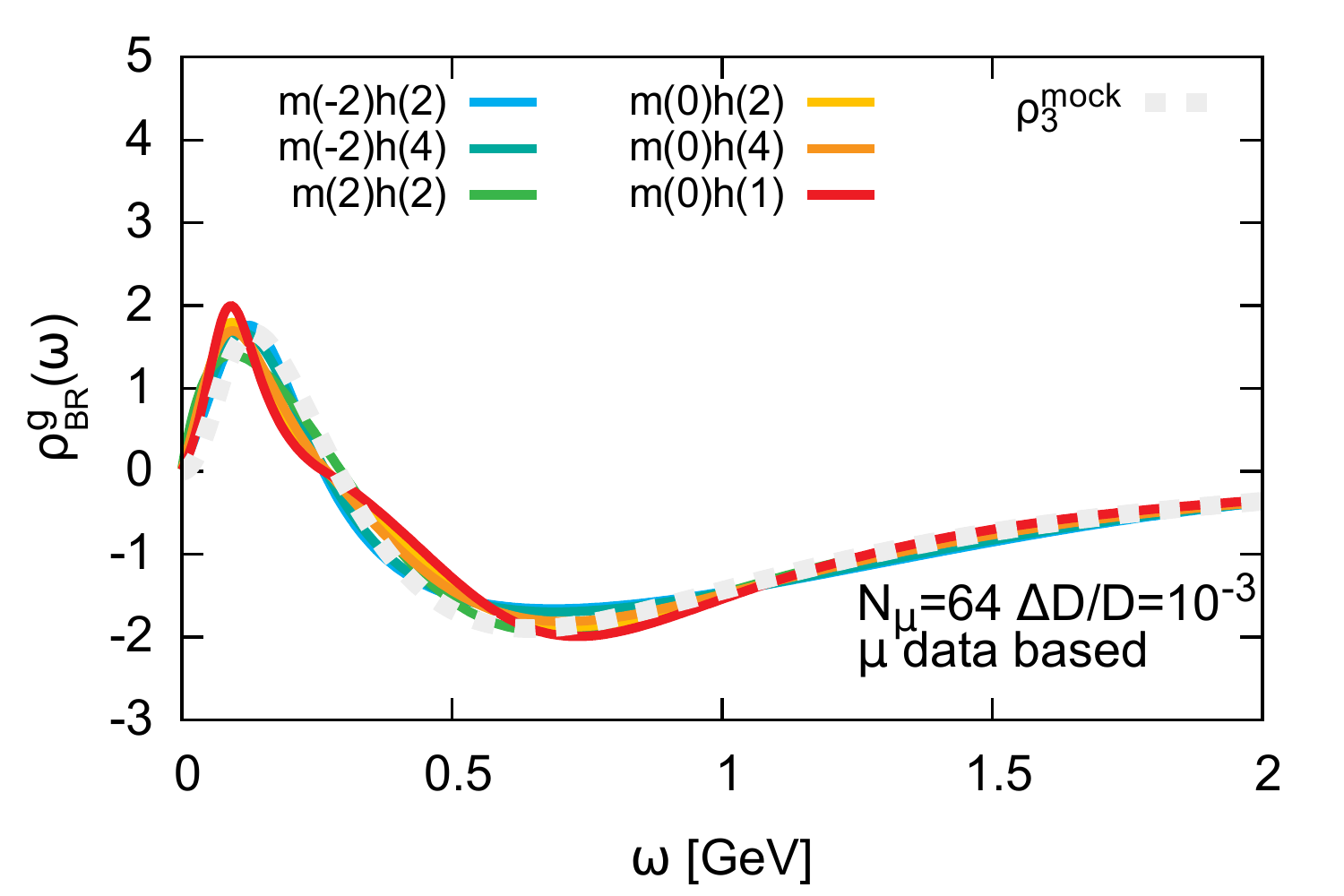}
\includegraphics[scale=0.5]{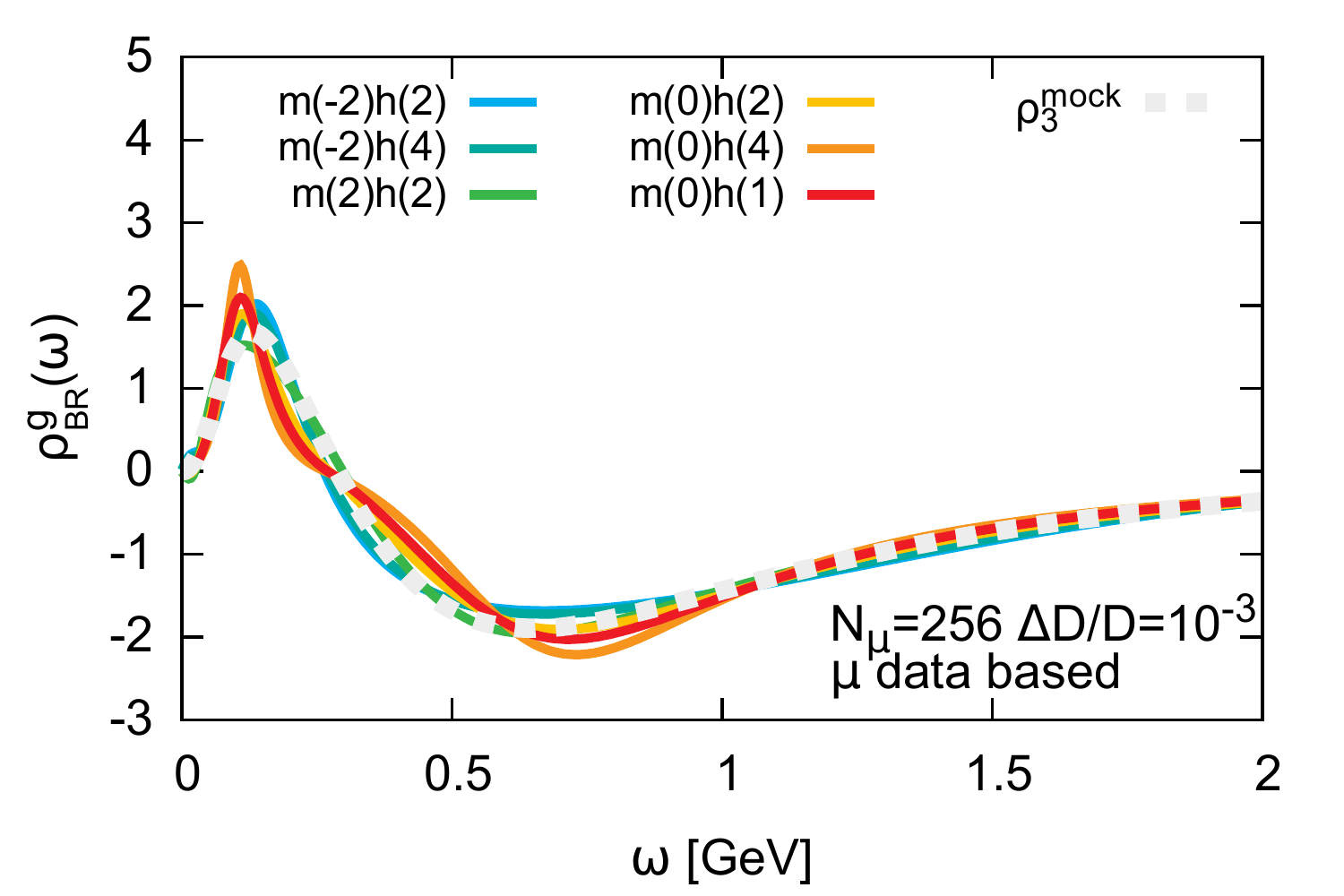}
\caption{Comparison of the very low frequency regime of the reconstructions (colored solid) for the mock spectrum (gray dashed) with exponential tail ($\rho^{\rm mock}_3$) at $\Delta D/D=10^{-3}$ for $N_\mu=64$ (left) and $N_\mu=256$ (right). We find that the reconstruction at $N_\mu=64$ already shows very robust behavior and the increase to $N_\mu=256$ appears unnecessary.}\label{Fig:ALG5KAELLENRecN64N256DefMod}
\end{figure*}

We also start here by presenting first reconstructions using the original BR method on shifted correlator data, using the same ten shift functions $\sigma$ as before. The outcome for $N_\mu=64$ is shown in Fig.~\ref{Fig:ALG5KAELLENRecN64}. The results are consistently at least as good as those obtained from the Euclidean data based on a kernel corresponding to $T=0.1$GeV. Hence they are much more accurate than the previous reconstructions at $T=1$GeV. In particular the dependence of the reconstructed peak position and width on the used shift function is weaker when using imaginary frequency data. Nevertheless we find that the use of shift functions may still lead to the occurrence of significant ringing e.g.\ for $\rho^{\rm mock}_1$ at larger frequencies.

The most important difference to the Euclidean based reconstructions lies in fact that increasing the number of datapoints here allows us to much more efficiently approach the "Bayesian continuum limit", as can be seen from the comparison in Fig.~\ref{Fig:ALG5KAELLENRecN64N256}. Quadrupling the number of datapoints from $N_\mu=64$ to $N_\mu=256$ significantly reduces the dependence of the reconstruction on the shift functions used. 

Once the uncertainty from the shift functions is under control, we may ask what is the residual dependence on the choice of default model, which we investigated by modifying the default model in two distinct ways. On the one hand we simply scale the shift function by a factor $\gamma=0.25,0.5,1,2,4$ before assigning it as default model $m=\gamma \sigma$, on the other hand we change the default model to a constant with either $m=15$ or $m=5$. We find that rescaling the default model away from the shift function $\gamma\neq 1$ leads to the appearance of very strong ringing artifacts, which distort not only the high but also the low frequency regime, in which the relevant peak structures reside.  While $m=\sigma$ is the most natural choice, the residual influence of the default model even at $N_\tau=256$ and $\Delta D/D=10^{-3}$ must be taken seriously, as the agreement between reconstruction and mock spectrum in the right panel of Fig.~\ref{Fig:ALG5KAELLENRecN64N256} can be an accident of our choice of setup and may not persist if a more realistic spectrum is considered.

In preparation of the investigation of actual gluon spectra, we learn that one should anticipate to provide around $N_\mu=256$ imaginary frequency datapoints with precision of around $\Delta D/D=10^{-3}$ to diminish the influence of the shift function to a level, where only the residual dependence on the choice of default model needs to be considered.

\subsubsection{Generalized BR method (I)}

Last but not least we investigate how the generalized BR method fares when deployed on non-shifted imaginary frequency data. Again we will show here nine reconstructions each, corresponding to different combinations of the default model $m$ and the function $h$. Fig.~\ref{Fig:ALG7KAELLENRecM1N64} contains the results for $N_\mu=64$ and we can see that decreasing the errorbars to $\Delta D/D=10^{-3}$ already leads to a very good quantitative reproduction of the mock spectrum with only a mild dependence on the default model.

The generalized BR method does not require the specification of a shift function and hence avoids $\sigma$ as a source of uncertainty. This in turn means that compared to the shift approach we may obtain quantitatively robust results already with a smaller number of imaginary frequency datapoints, as shown in Fig.~\ref{Fig:ALG5KAELLENRecN64N256DefMod}. Note that in the generalized BR reconstruction, once we are at $\Delta D/D\leq 10^{-2}$, we do not find artificial ringing around the lowest lying peak. I.e. below $\omega=2GeV$ there is a single positive peak and a single negative trough present in the reconstruction.

\begin{figure*}[t!]
\includegraphics[scale=0.38]{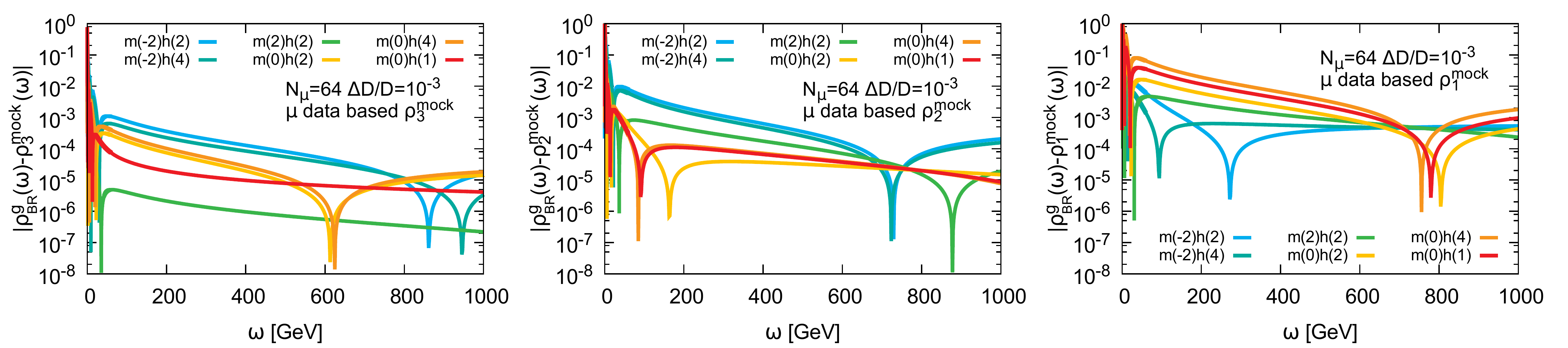}
\caption{Approach of the reconstructed spectra to the asymptotic tail behavior at high frequencies. We show the difference $|\rho^g_{BR}-\rho^{\rm mock}|$ at $\Delta D/D=10^{-3}$ for $\rho^{\rm mock}_3$ (left), $\rho^{\rm mock}_2$ (center) and $\rho^{\rm mock}_1$ (right) each for six different choices of the default model parameters. As expected for most choices of the default model, the approach improves the faster the tail dampens. Nevertheless artificial oscillations around the correct result, manifest on the logarithmic scale as sharp dips, do persist up to high frequencies, even for the exponential tail.}\label{Fig:ALG7KAELLENRecAsymptotic}
\end{figure*}

While ringing at low frequencies appears well under control, it can still affect the approach of the reconstruction to the asymptotic tail behavior encoded in the mock spectra. In Fig.~\ref{Fig:ALG7KAELLENRecAsymptotic} we show the difference $|\rho^g_{BR}-\rho^{\rm mock}|$ between the mock spectrum and the reconstructions from $N_\mu=64$ datapoints at $\Delta D/D=10^{-3}$. As expected, the faster the tail in the mock spectrum damps, the better the reconstruction is able to capture it. Far in the UV, at $\omega=1000$GeV the exponential tail reconstruction shows a deviation of $\lesssim10^{-5}$, while for the quadratic tail we have deviations between $10^{-4}-10^{-5}$. In case of the unphysical $-1/x$ tail, the deviation varies around $10^{-3}$. The presence of sharp dips in the logarithmic plot indicates that artificial oscillations around the correct answer do occur, albeit with diminishing amplitude, for most choices of the default models up to very high frequencies. These artifacts however do not appear to significantly impede the reconstruction of the low lying spectral features as discussed in regard to Fig.~\ref{Fig:ALG7KAELLENRecM1N64} and Fig.~\ref{Fig:ALG5KAELLENRecN64N256DefMod}.

The application of the generalized BR method thus bodes well for the study of gluon spectra from functional QCD methods, as it appears to provide an efficient prescription to already extract non-positive definite spectra using a relatively small number of datapoints $N_\mu=64$ at $\Delta D/D=10^{-3}$.

\section{Conclusion}
\label{sec:Conclusion}

We have introduced a novel Bayesian approach to the reconstruction of general spectral functions, which contain both positive and negative contributions. It is closely related to the recently developed Bayesian reconstruction (BR) method and based on a similar set of axioms. In particular it retains the quality that it imprints the prior information in the default model in a very weak fashion while still leading to a unique reconstruction. 

We were lead to a different functional form of the prior $S^g_{\rm BR}$ compared to the original BR method, since for general spectra $\rho=0$ and $m=0$ are admissible values and we hence replaced $\rho/m$ by $|\rho-m|/h+1$ as measure of deviation between spectrum and default model. To maintain scale invariance, i.e.\ the independence of the end result from the units of $\rho$, we needed to introduce an additional function $h$, which may be interpreted as our confidence in the values of $m$.

In anticipation of the study of gluon spectra from Landau gauge correlators computed in lattice QCD \cite{Ilgenfritz:2017kkp} and from functional methods in QCD \cite{Pawlowski:2016inprep}, we performed an extensive set of mock data tests, comparing our novel direct approach to a standard strategy based on shift functions. Since in functional computations correlators can be evaluated both in Euclidean time and imaginary frequency space, we checked how the form of the kernel influences the success of the spectral reconstruction.

We found a striking manifestation of the exponential information loss induced by the Euclidean kernel, when reconstructing from imaginary time data. In case of the shift method, even for $\Delta D/D=10^{-3}$ and $N_\tau=256$ the results remained strongly dependent on the choice of shift function and increasing the number of datapoints lead to virtually no visible improvements. The higher the temperature, the more severe this issue becomes. 

Since this loss of information is independent from the method used for spectral reconstruction also the generalized BR method was found to struggle with these reconstructions. While it avoids the additional uncertainty introduced by shift functions, its results may depend significantly on the choice of default models $m$ and $h$. Also here increasing the number of datapoints does not lead to a visible improvement.

The situation is very different when instead using imaginary frequency data. The method based on shifted data already shows a much weaker dependence on the choice of shift function for $N_\mu=64$ than in the Euclidean case and increasing the number of datapoints actually leads to a significant improvement of the accuracy and precision in the reconstruction. 

Nevertheless the shift method carries an inherent uncertainty due to the choice of shift function $\sigma$, which to be brought under control requires us to use at least $N_\mu=256$ points. In contrast the novel generalized BR method avoids this additional source of systematic error and thus shows a robust quantitative reconstruction of spectral features already for $N_\mu=64$ with only a very mild dependence on the variation of default models $m$ and $h$. 

We therefore are confident that the generalized BR method, when applied to imaginary frequency data, will provide a useful and efficient tool for deconvolution problems in quantum field theory and beyond. In light of a recently proposed approach \cite{Pawlowski:2016} to simulating thermal quantum fields directly in imaginary frequencies, it promises to benefit not only reconstructions within continuum approaches to QCD but also those based on correlators from numerical simulations.

The author thanks Y. Burnier for fruitful collaboration on the derivation and application of the original BR method. Collaboration on related projects with J.~M.~Pawlowski are gladly acknowledged. This work is part of and supported by the DFG Collaborative Research Centre "SFB 1225 (ISOQUANT)".

\end{document}